\documentclass[A4paper]{article}
\usepackage{amsmath}
\usepackage{amssymb}
\usepackage{graphicx,psfrag,epsf}
\usepackage{enumerate}
\usepackage{natbib}
\usepackage{url} 
\usepackage{fullpage}
\usepackage{algorithm}
\usepackage{algorithmic}

\newcommand{\blind}{1}

\renewcommand\footnotemark{}

\newcommand{\bt}{\mathbf{t}}
\newcommand{\by}{\mathbf{y}}
\newcommand{\bv}{\mathbf{v}}

\newcommand{\bb}{\mathbf{b}}
\newcommand{\bm}{\mathbf{m}}
\newcommand{\be}{\mathbf{e}}

\newcommand{\bbeta}{\boldsymbol{\beta}}

\newcommand{\bxi}{\boldsymbol{\xi}}
\newcommand{\bmu}{\boldsymbol{\mu}}
\newcommand{\btheta}{\boldsymbol{\theta}}
\newcommand{\bdelta}{\boldsymbol{\delta}}
\newcommand{\btau}{\boldsymbol{\tau}}

\newcommand{\bgamma}{\boldsymbol{\gamma}}
\newcommand{\bseta}{\boldsymbol{\eta}}

\newcommand{\bzero}{\mathbf{0}}

\newtheorem{theorem}{Theorem}[section]

\newtheorem{lemma}[theorem]{Lemma}

\def\T{{ \mathrm{\scriptscriptstyle T} }}

\begin{document}

\def\spacingset#1{\renewcommand{\baselinestretch}%
{#1}\small\normalsize} \spacingset{1}


\if1\blind
{
  \title{\bf \Large Bayesian decision-theoretic design of experiments under an alternative model
  \vspace{0cm}
  }
  \author{\large Antony M. Overstall\hspace{.2cm}\\
    \large Southampton Statistical Sciences Research Institute,\\ \large University of Southampton,\\ \large Southampton, UK\\ \large (A.M.Overstall@soton.ac.uk) \\[1ex]
    \large James M. McGree\\
    \large School of Mathematical Sciences,\\ \large Queensland University of Technology,\\ \large Brisbane, Australia \\ \large (james.mcgree@qut.edu.au)}
    \date{\vspace{-1.2cm}}
  \maketitle
} \fi

\if0\blind
{
  \bigskip
  \bigskip
  \bigskip
  \begin{center}
    {\LARGE\bf Title}
\end{center}
  \medskip
} \fi

\bigskip
\begin{abstract}
Traditionally Bayesian decision-theoretic design of experiments proceeds by choosing a design to minimise expectation of a given loss function over the space of all designs. The loss function encapsulates the aim of the experiment, and the expectation is taken with respect to the joint distribution of all unknown quantities implied by the statistical model that will be fitted to observed responses. In this paper, an extended framework is proposed whereby the expectation of the loss is taken with respect to a joint distribution implied by an alternative statistical model. Motivation for this includes promoting robustness, ensuring computational feasibility and for allowing realistic prior specification when deriving a design. To aid in exploring the new framework, an asymptotic approximation to the expected loss under an alternative model is derived, and the properties of different loss functions are established.  The framework is then demonstrated on a linear regression versus full-treatment model scenario, on estimating parameters of a non-linear model under model discrepancy and a cubic spline model under an unknown number of basis functions.
\end{abstract}
\noindent%

\section{Introduction} \label{sec:intro}

The Bayesian decision-theoretic approach \citep{chalonerverdinelli1995} is a natural framework to plan experiments in many fields of science and engineering. It starts with specification of a loss function representing the aim of the experiment. A Bayesian design then minimises the expectation of the loss over the space of all possible designs where expectation is with respect to a probability distribution over all unknown quantities implied by the statistical model that will be fitted upon observation of the experimental responses. 

The approach is scientifically appealing since the precise aim of the experiment is encoded by the loss function and, through specification of a probability distribution over all unknown quantities, pre-experimental (lack of) knowledge is incorporated. In practice, finding a Bayesian design by minimising the expected loss is a challenging computational problem \citep{ryanetal2016}. The expected loss is typically analytically intractable and the space of all designs potentially high-dimensional. However, recently there has been significant progress in the development of novel computational methodology for finding Bayesian designs originating in diverse fields of study \citep[e.g.][and references therein]{long_etal_2013, ryanetal2016, foster2019, beck_etal_2020}.

In this paper, we propose an extended framework for Bayesian decision-theoretic design of experiments. This is achieved by defining the expected loss by taking expectation with respect to a probability distribution implied by an alternative statistical model (termed a designer model).  \cite{EtzioniKadane1993} considered the case where the fitted and designer prior distributions were different but the joint distributions for the responses were identical. We extend this by considering the case where both the prior \emph{and} distribution for responses can be different. There are several reasons why a design may be sought under a different model to the fitted model. For example, the fitted model may emulate a partially observed process, but by considering the whole data-generating process, robustness is introduced into the design procedure. Assuming a simpler fitted model can also be used to induce robustness to model misspecification \citep{roshan_2009} or may be necessary to ensure computational feasibility of finding Bayesian designs \citep{cmryan2016}. Conversely, reliable prior specification can be assisted by the designer model being simpler than the fitted model.  Accordingly, we provide the framework and supporting methodology to enable Bayesian design within such settings, and apply these methods to typical design problems found in the literature.

The outline of the paper is as follows. In Section \ref{sec:meth}, the extended framework for Bayesian design of experiments under an alternative model is described and justified, and the choice of loss function considered. In Section~\ref{sec:explore}, an asymptotic approximation to the expected loss under the extended framework is developed to explore the consequences of designing under an alternative model. Under the traditional approach, designs found under a loss, and its posterior expectation, are equivalent. This is no longer the case under the proposed extended framework. We consider the choice between these pairs of loss functions and show that there is essentially a bias-variance trade off to be made. To demonstrate the extended framework, three examples are considered in Section~\ref{sec:examples} and Section~SM6 in the Supplementary Material. The first considers a linear regression versus full-treatment model scenario where application of the extended framework results in objective functions sharing properties of classical design criteria that promote increased replication \citep{gilmourtrinca2012}. The second considers estimating parameters of non-linear models under model discrepancy, and finds that the resulting designs resemble a compromise of designs found under the traditional Bayesian decision-theoretic framework and space-filling designs. Finally, the third (in the Supplementary Material), considers estimation of the parameters of a model formed using cubic splines, with an unknown number of basis functions.

The Supplementary Material includes proofs and derivations of the results, and the cubic spline model example.
 
\section{Bayesian design under an alternative model} \label{sec:meth}
 
%
\subsection{Internal and external expected loss} \label{sec:exploss}

Suppose the ultimate aim of the experiment is to learn the relationship between $k$ controllable variables and a response. The experiment consists of $n$ runs where, for $i=1,\dots,n$, the $i$th run involves specifying a treatment-level combination, i.e. values for the $k$ controllable variables $\Delta_i = \left(\Delta_{i1}, \dots, \Delta_{ik} \right)^\T  $, and subsequent observation of a response $y_i$. Let $\Delta$ denote the design; the $n \times k$ matrix with $i$th row given by $\Delta_i^\T  $ and let $\by = \left(y_1,\dots,y_n\right)^\T  $ denote the $n \times 1$ vector of observed responses. 
 
On the completion of the experiment, a statistical model is assumed describing the relationship between the controllable variables and response. That is, $\by$ is assumed to be a realisation from a multivariate probability distribution with density/mass function $\pi(\by|\bbeta, F,\Delta)$ completely known up to a $p \times 1$ vector of unknown parameters $\bbeta$ with prior density $\pi(\bbeta|F,\Delta)$. We refer to $\pi(\by|\bbeta, F,\Delta)$ and $\pi(\bbeta|F,\Delta)$ as the \emph{fitted likelihood} and \emph{prior}, respectively, and collectively as the \emph{fitted model}, where the conditioning on $F$ is to make it clear that this is under the fitted model. Using Bayes' theorem, the \emph{fitted posterior} distribution is $\pi(\bbeta|\by, F,\Delta) \propto \pi(\by|\bbeta, F,\Delta) \pi(\bbeta \vert F,\Delta)$. Note that $\bbeta$ refers to parameters of interest. There may be additional nuisance parameters present but these have been marginalised, with respect to a prior distribution, to obtain $\pi(\by|\bbeta, F,\Delta)$.

Since, apart from the values of $\bbeta$, the fitted model completely specifies the relationship between the controllable variables and response, the experimental aim reduces to determining the value of $\bbeta$. This aim can manifest itself in various different ways. Bayesian decision-theoretic design of experiments starts by encoding the exact aim in a loss function, $\lambda(\bbeta, \by, F,\Delta)$, giving the loss of estimating $\bbeta$ via the fitted posterior distribution conditional on $\by$. The choice of loss function is considered in more detail in Section~\ref{sec:losses}. 

Prior to experimentation, the aim is to specify $\Delta$ to best estimate $\bbeta$ as measured by the loss function. Traditionally, a Bayesian design minimises the expected loss over the space of all designs, where expectation is with respect to the joint distribution of $\bbeta$ and $\by$ as implied by the fitted model \citep[e.g.][]{lindley_1972}. Mathematically, the expected loss is 
\begin{equation}
L_{FF}(\Delta) = \mathrm{E}_{\bbeta, \by \vert F,\Delta} \left[ \lambda(\bbeta, \by, F,\Delta) \right]
\label{eqn:0}
\end{equation}
and referred to, in this paper, as the \emph{internal expected loss} under the fitted model. Unless otherwise specified; if referring to internal expected loss, it is under the fitted model.

Instead, consider minimising the expectation of the loss with respect to an alternative model: the \emph{designer model}. Here we suppose that $\by$ is a realisation of a probability distribution with density/mass $\pi(\by|\btheta, D,\Delta)$, where $\btheta$ is a $p_{\theta} \times 1$ vector of parameters with prior density $\pi(\btheta|D,\Delta)$. We refer to $\pi(\by|\btheta, D,\Delta)$ and $\pi(\btheta|D,\Delta)$ as the \emph{designer likelihood} and \emph{prior}, respectively, and the conditioning on $D$ is to make it clear that this is under the designer model. Similar to the fitted model, any additional nuisance parameters have been marginalised to obtain $\pi(\by|\btheta, D,\Delta)$. 

Let $\bbeta = \left(\bgamma^\T, \btheta^\T\right)^\T$, where $\bgamma$ is a $p_{\gamma} \times 1$ vector of parameters only present in the fitted model and $\btheta$ are parameters common to both models. If the fitted and designer models share no common parameters, i.e. $\bbeta = \bgamma$, then the two models are referred to as \emph{disjoint}. Conversely, if all parameters of interest are common to both models, i.e. $\bbeta = \btheta$, then the models are referred to as \emph{compatible}. As such, prior to marginalisation, the fitted model can be nested within the designer model, vice-versa, or no nesting may be present.

A design found under the proposed extended framework is given by minimising the \emph{external expected loss} defined as
\begin{equation}
L_{DF}(\Delta) = \mathrm{E}_{\btheta, \by \vert D,\Delta} \left\{ \mathrm{E}_{\bgamma| \btheta, \by, F,\Delta} \left[ \lambda(\bbeta, \by, F,\Delta)\right] \right\}.
\label{eqn:designer}
\end{equation}
The external expected loss is formed by assuming that once responses are observed, we proceed as if the fitted model represents the true data-generating process, i.e. under the M-closed paradigm \citep[][page 384]{bernardo_smith_1994}. Initially, expectation is taken with respect to the fitted posterior distribution of $\bgamma$ conditional on $\btheta$. This is because these parameters are only present in the fitted model (if the models are non-compatible). Then expectation is taken with respect to the remaining unknowns under the designer model.

Note that if the fitted and the designer models are the same, then the external and internal expected loss functions are equal, i.e. $L_{DF}(\Delta) = L_{FF}(\Delta)$.

There are several reasons to consider finding a design under an alternative model. These can be separated into two cases: when the designer model is simpler than the fitted model and the vice-versa. In Section~\ref{sec:examples} and Section~SM6 in the Supplementary Material we provide examples of both cases.

First consider the case where the fitted model actually represents the, \emph{a-priori}, current best representation of the data-generating process, and the designer model is actually a proxy simplification. This can occur when the fitted model is complex with a large number of parameters and it is necessary to use a fitted prior distribution representing weak prior information. It is known \citep[e.g.][]{ryanetal2016} that in these cases, the vaguery of the fitted prior distribution combined with the large number of parameters leads to the internal expected loss being relatively flat and, therefore, non-trivial to (numerically) minimise. However it may be possible to elicit a prior distribution for a simpler designer model and use this to take expectation of the loss function formed under the fitted model.

Now consider the case where the designer model actually represents the best representation of the data-generating process and the fitted model is a simplification. In this case, it initially would make sense to find a design by minimising the internal expected loss under the designer model, i.e.
\begin{equation}
L_{DD}(\Delta) = \mathrm{E}_{\btheta, \by \vert D,\Delta} \left[ \lambda(\btheta, \by, D,\Delta) \right].
\label{eqn:dd}
\end{equation}
For certain models it may not be possible to represent the experimental aim by a suitable loss function, $\lambda(\btheta, \by, D,\Delta)$, depending on the designer posterior distribution. Instead, an alternative fitted model is used for which it is possible to form a suitable loss function. However it is prudent to take expectation under the original designer model since this is the best representation of the data-generating process. For example, suppose the designer model is a full-treatment model \citep[e.g.][]{gilmourtrinca2012} with each unique treatment-level combination permitting a unique mean response. This represents a plausible data-generating process but for convenience and interpretability a regression model (positing a relationship between the numerical values of the controllable variables and mean response) is usually assumed as the fitted model.

Alternatively, suppose the designer model is based on scientific reasoning through a mechanistic or phenomenological model. Again it would make sense to find the design via the internal expected loss under the designer model (\ref{eqn:dd}). Assuming the fitted and designer models are compatible, consider Monte Carlo approximations to the external and internal expected loss functions given by
$\hat{L}_{DF}(\Delta) = B^{-1} \sum_{b=1}^B \lambda\left(\btheta_b, \by_b, F,\Delta \right)$ and $\hat{L}_{DD}(\Delta) = B^{-1} \sum_{b=1}^B \lambda\left(\btheta_b, \by_b, D,\Delta \right)$, respectively, where $\left\{ \btheta_b, \by_b \right\}_{b=1}^B$ is a sample generated from the joint distribution of $\btheta$ and $\by$ under the designer model. There are models, e.g. so-called intractable likelihood models, for which it is straightforward to generate such a sample but extremely difficult to evaluate the designer posterior distribution and hence the loss function $\lambda\left(\btheta_b, \by_b, D ,\Delta\right)$, for $b=1,\dots,B$. However by forming an approximating fitted model it becomes computationally feasible to find a Bayesian design via the external expected loss \citep[e.g., see the indirect inference approach of ][]{cmryan2016}. In Section~\ref{sec:losses}, we show, under a certain class of loss function, that the external expected loss is an upper bound on the internal expected loss under the designer model, providing formal justification for using such approaches.

\subsection{Loss functions} \label{sec:losses}

In Section~\ref{sec:exploss}, the loss function is given informally by $\lambda(\bbeta, \by, F,\Delta)$, stating that it gives the loss of estimating $\bbeta$ via the fitted posterior (conditional on $\by$). We now consider the specification of the loss and the implications of such a choice when designing an experiment under an alternative model. For this, we consider two exemplar loss functions as defined in Table \ref{tab:loss} (left-hand column); self-information and squared error. Note that $\Vert \bv \Vert_{2,A}^2 = \bv^\T   A \bv$ and $I_p$ is the $p \times p$ identity matrix. These loss functions have been chosen since under a normal linear model with non-informative prior distributions, the internal expected loss reduces to the objective function under D- and A-optimality, respectively; two widely-studied classical optimal design criteria. 

\begin{table}
\begin{center}
\caption{Exemplar generator and composite loss functions. \label{tab:loss}}
\begin{tabular}{llll} \hline \hline
\multicolumn{2}{c}{Generator} & \multicolumn{2}{c}{Composite} \\ \hline
Name & Expression for & Name & Expression for  \\ 
 & $\lambda(\bbeta, \by,F,\Delta)$ & & $\lambda(\by,F,\Delta)$ \\ \hline
 \\
Self- & $\lambda_{SI}(\bbeta, \by,F,\Delta) = $  & Entropy & $\lambda_{E}(\by,F,\Delta) = $\\ 
information & $- \log \pi(\bbeta | \by, F,\Delta)$ & (E) & $\mathrm{E}_{\bbeta | \by, F,\Delta}\left[- \log \pi(\bbeta | \by, F,\Delta)\right]$\\ 
(SI) & & & \\ \hline
\\
Squared & $\lambda_{SE}(\bbeta, \by,F,\Delta) = $ & Trace & $\lambda_{TV}(\by,F,\Delta) = $\\ 
error & $\Vert \bbeta - \mathrm{E}_{\bbeta \vert \by, F,\Delta}\left(\bbeta \right) \Vert^2_{2,I_p}$& variance & $\mathrm{tr}\left[\mathrm{var}_{\bbeta | \by, F,\Delta}\left(\bbeta\right) \right]$\\ 
(SE) & & (TV) & \\ \hline \hline
\end{tabular}
\end{center}
\end{table}

\subsubsection{Optimal loss functions}

First consider estimating $\bbeta$ based on observed responses $\by$. Suppose the loss for this aim is $\lambda_0(\bbeta, \bb)$ where $\bb \in \mathcal{B}$ is an action for estimating $\bbeta$ and $\mathcal{B}$ is the action space (e.g. $\bb$ is a point estimate of $\bbeta$ in $\mathcal{B} = \mathbb{R}^p$). The optimal action minimises the fitted posterior expectation of $\lambda_0(\bbeta, \bb)$, i.e. $\hat{\bb}_F = \mathrm{arg} \min_{\bb \in \mathcal{B}} \mathrm{E}_{\bbeta \vert \by, F,\Delta} \left[ \lambda_0(\bbeta, \bb)\right]$.

The loss function, as written in Section~\ref{sec:exploss}, is given by $\lambda(\bbeta, \by, F,\Delta) = \lambda_0(\bbeta, \hat{\bb}_F)$ and is optimal under the fitted posterior distribution. In this paper, only optimal loss functions are considered. Indeed, the self-information and squared error loss functions are optimal under logarithmic scoring and $\lambda_0(\bbeta, \bb) = \Vert \bbeta - \bb \Vert_{2,I_p}^2$, respectively. However, the formulation in Section~\ref{sec:exploss} allows consideration of sub-optimal loss functions, if deemed appropriate, e.g. 
$\lambda(\bbeta, \by, F,\Delta) = \Vert \bbeta - \mathrm{M}_{\bbeta \vert \by, F,\Delta} \left( \bbeta \right) \Vert_{2,I_p}^2$, where $\mathrm{M}_{\bbeta \vert \by, F,\Delta} \left( \bbeta \right)$ is the fitted posterior mode of $\bbeta$.


Under optimal loss functions, the following result giving an inequality for the external expected loss in relation to the internal expected loss under the designer model. The proof is given in Section~SM1 of the Supplementary Material.

\begin{lemma} \label{lem:equality}
Suppose fitted and designer models are compatible, i.e. $\bbeta = \btheta$, with $\lambda(\bbeta, \by, F,\Delta) = \lambda_0(\bbeta, \hat{\bb}_F)$ and $\lambda(\bbeta, \by, D,\Delta) = \lambda_0(\bbeta, \hat{\bb}_D)$, then $L_{DF}(\Delta) \ge L_{DD}(\Delta)$, where $L_{DD}(\Delta)$ is the internal expected loss under the designer model given by (\ref{eqn:dd}).
\end{lemma}

%

In Lemma~\ref{lem:equality}, the difference between the external expected loss and the internal expected loss under the designer model is due to the loss of information caused by the fitted model being different to the model under which expectation is taken. Note that Lemma 2.1 only requires the compatible fitted and designer models to be different, not that the designer model is more complex.

By finding the design that minimises the external expected loss, we are minimising an upper bound on the internal expected loss under the designer model. A special case of Lemma~\ref{lem:equality} for the self-information loss is called the Barber-Agakov bound \citep{barberagakov2003} and has been used by \citet{foster2019} to approximately find Bayesian designs. Lemma~\ref{lem:equality} now justifies the use of this approach to other loss functions as has previously been pursued for intractable likelihood models (for example, \citealt{cmryan2016}, as discussed in Section~\ref{sec:exploss}).

Lemma~\ref{lem:equality} is uninformative on the relationship between the external expected loss and the internal expected loss under the fitted model (\ref{eqn:0}).

\subsubsection{Generator and composite loss functions}

There are intuitive choices of loss function given as scalar summaries of the fitted posterior distribution, independent of $\bbeta$, e.g. $\lambda(\by, F,\Delta) = \vert\mathrm{var}(\bbeta \vert \by, F,\Delta) \vert$, where $\vert \cdot \vert$ denotes the determinant of a square matrix. In some cases, these are given as the fitted posterior expectation of a loss function $\lambda_G(\bbeta, \by, F,\Delta)$, i.e. $\lambda_C(\by, F,\Delta) = \mathrm{E}_{\bbeta \vert \by, F,\Delta} \left[ \lambda_G(\bbeta, \by, F,\Delta)\right]$. For such pairs, we term $\lambda_C(\by, F,\Delta)$ as composite and $\lambda_G(\bbeta, \by, F,\Delta)$ as generator. The right-hand column of Table~\ref{tab:loss} shows the entropy and trace variance loss functions which are composite to the generator loss functions of self-information and squared error, respectively, in the left-hand column.

The internal expected composite loss is equal to the corresponding internal expected generator loss. Apart from when fitted and designer models are disjoint, i.e. $\bbeta  = \bgamma$, this equivalence is not necessarily true when considering the external expected loss. This leads to a question on how to choose between a pair of loss functions equivalent under internal expected loss, e.g. how should we choose between self-information and entropy loss functions? In the next section, to answer this question, we develop an asymptotic approximation to the external expected loss. This is used to explore differences in behaviour between the exemplar generator and composite loss functions shown in Table~\ref{tab:loss}. We find that a composite loss solely focuses on fitted posterior precision. On the other hand, a generator loss also incorporates point estimation of common parameters $\btheta$. This is essentially a bias-variance trade-off. Considering this trade-off is unnecessary under internal expected loss because the fitted model is, in a sense, ``unbiased'' as it is identical to the designer model.

\section{Understanding the external expected loss} \label{sec:explore}

In this section, an asymptotic approximation to the external expected loss is derived.  This approximation is used to provide insight into properties of exemplar loss functions, in particular, to investigate the difference between external expected generator and composite loss functions.

\subsection{Asymptotic approximation of external expected loss} \label{sec:general}

In general, the external expected loss (\ref{eqn:designer}) is not available in closed form and use of an approximation is therefore necessary. Recent developments in computational methodology for approximating (and minimising) the expected loss (example references in Section~\ref{sec:intro}) can actually be used for this task with little modification. Indeed, in the examples in Section~\ref{sec:nlm} and Section~SM6 in the Supplementary Material we describe Monte Carlo approximations to the external expected loss function.

However, in this section, to gain understanding of the behaviour of the external expected loss, we develop an asymptotic approximation. The resulting approximations to the external expected loss under the exemplar loss functions in Table~\ref{tab:loss} are derived and compared to the corresponding expressions under the internal expected loss. 

First, some preliminary definitions are required. Let $\tilde{\bbeta}= \left( \tilde{\bgamma}^\T,\tilde{\btheta}^\T\right)^\T$ be the $p \times 1$ vector of parameter values for $\bbeta$ that minimise the Kullback-Leibler divergence between the designer and fitted likelihoods for a given value of $\btheta$ under the designer model, i.e. $\tilde{\bbeta}$ maximises
\begin{equation}
d(\bt) = \mathrm{E}_{\by\vert\btheta,D,\Delta} \left[ \log \pi (\by\vert \bt, F,\Delta) \right].
\label{eqn:dfunc}
\end{equation}
Define $\mathcal{I}$ and $\tilde{\mathcal{I}}$ to be 
$$\mathrm{E}_{\by \vert \btheta, D,\Delta} \left[ - \frac{\partial^2 \log \pi (\by\vert \bt, F,\Delta)}{\partial \bt \partial \bt^\T } \right]$$
evaluated at $\bt = \btheta$ and $\bt = \tilde{\btheta}$, respectively. Similarly, define $\tilde{\mathcal{J}}$ to be 
$$\mathrm{E}_{\by \vert \btheta, D,\Delta} \left[ \frac{\partial \log \pi (\by\vert \bt, F,\Delta)}{\partial \bt} \frac{\partial \log \pi (\by\vert \bt, F,\Delta)}{\partial \bt^\T } \right], $$
evaluated at $\bt = \tilde{\btheta}$, respectively. Let $\tilde{\mathcal{K}} = \tilde{\mathcal{I}}^{-1} \tilde{\mathcal{J}}\tilde{\mathcal{I}}^{-1}$ with $\tilde{\mathcal{I}}$ and $\tilde{\mathcal{K}}$ partitioned as
$$\tilde{\mathcal{I}} = \left[ \begin{array}{cc}
\tilde{\mathcal{I}}_{\gamma \gamma} & \tilde{\mathcal{I}}_{\gamma\theta}\\
\tilde{\mathcal{I}}_{\theta \gamma} & \tilde{\mathcal{I}}_{\theta \theta}
\end{array} \right], \qquad \qquad \tilde{\mathcal{K}} = \left[ \begin{array}{cc}
\tilde{\mathcal{K}}_{\gamma \gamma} & \tilde{\mathcal{K}}_{\gamma\theta}\\
\tilde{\mathcal{K}}_{\theta \gamma} & \tilde{\mathcal{K}}_{\theta \theta}
\end{array} \right]$$
where $\tilde{\mathcal{I}}_{\gamma \gamma}$ is the $p_\gamma \times p_\gamma$ sub-matrix corresponding to $\bgamma$ with similar definitions for $\tilde{\mathcal{I}}_{\theta \theta}$, $\tilde{\mathcal{I}}_{\theta \gamma}= \tilde{\mathcal{I}}_{\gamma\theta}^\T $ and the sub-matrices of $\tilde{\mathcal{K}}$. 

Under regularity conditions (see Section~SM2 in the Supplementary Material), an asymptotic approximation \citep{kleijn_vandervaart_2012} to the fitted posterior distribution of $\bbeta$ is given by
\begin{equation}
\mathrm{N}\left(\hat{\bbeta}, \tilde{\mathcal{I}}^{-1}\right),
\label{eqn:berger}
\end{equation}
where $\hat{\bbeta} = \left( \hat{\bgamma}^\T,\hat{\btheta}^\T\right)^\T$ is the maximum likelihood estimate of $\bbeta$ under the fitted model.

The loss function is approximated by replacing fitted posterior quantities by the corresponding approximated fitted posterior quantities under (\ref{eqn:berger}). Such an approximation is denoted $\lambda^*(\bbeta, \hat{\bbeta},\Delta)$. For example, the self-information loss is approximated by
$$
\lambda^*_{SI}(\bbeta, \hat{\bbeta},\Delta) = N_{SI} - \frac{1}{2} \log \vert \tilde{\mathcal{I}} \vert + \frac{1}{2}\Vert \bbeta - \hat{\bbeta} \Vert^2_{2,\tilde{\mathcal{I}}},
$$
where $N_{SI} = p \log (2\pi)/2$ is a constant not depending on the design. 

To approximate the external expected loss (\ref{eqn:designer}), the parameters present only in the fitted model, $\bgamma$, are marginalised with respect to an approximate fitted posterior conditional on $\btheta$, induced from (\ref{eqn:berger}). Then expectation with respect to the distribution of $\by$ conditional on $\btheta$ under the designer model can be approximated using asymptotic results for inference under an alternative (or misspecified) model \citep{white1982}.

The result is now formally stated with proof given in Section~SM2 of the Supplementary Material.

\begin{theorem} \label{th:main}
Under conditions given in Section~SM2 of the Supplementary Material, an asymptotic approximation to the external expected loss is given by
\begin{equation}
L^*_{DF}(\Delta) = \mathrm{E}_{\btheta \vert D,\Delta} \left( \mathrm{E}_{\hat{\bbeta} \vert \btheta, D,\Delta} \left\{ \mathrm{E}_{\bgamma \vert \hat{\bbeta}, \btheta, F,\Delta} \left[ \lambda^*(\bbeta, \hat{\bbeta},\Delta) \right] \right\} \right)
\label{eqn:result}
\end{equation}
where the two inner expectations are taken with respect to
$$
\bgamma \vert \hat{\bbeta}, \btheta, F,\Delta \sim \mathrm{N} \left( \tilde{P}_1 \hat{\bbeta} - \tilde{P}_2 \btheta, \tilde{\mathcal{I}}_{\gamma \gamma}^{-1} \right), \qquad \mbox{and} \qquad
\hat{\bbeta} \vert \btheta, D,\Delta \sim \mathrm{N} \left( \tilde{\bbeta}, \tilde{\mathcal{K}}\right),
$$
with $\tilde{P}_1 = \left(I_{p_\gamma}, \tilde{P}_2 \right)$ and $\tilde{P}_2 = \tilde{\mathcal{I}}_{\gamma \gamma}^{-1} \tilde{\mathcal{I}}_{\gamma \theta}$.
\end{theorem}

In Theorem~\ref{th:main}, $\tilde{\mathcal{K}}$ has the sandwich form of the ``robust variance formula" \citep[e.g.][page 374]{pawitan2003}. In inferential settings, the values of $\tilde{\mathcal{I}}$ and $\tilde{\mathcal{J}}$ are estimated from observed responses, whereas, in design settings, the values follow from the assumption of a designer model.

Due to the tractability of the normal distribution, the two inner expectations in (\ref{eqn:result}) are available in closed form for many common loss functions, e.g. the exemplar loss functions in Table~\ref{tab:loss}. Then the approximate external expected loss is written $L^*_{DF}(\Delta) = \mathrm{E}_{\btheta \vert D,\Delta} \left[ \ell^*_{G}(\btheta,\Delta) \right]$; the designer prior expectation of a function $\ell^*_{G}(\btheta,\Delta)$, typically a function of $\tilde{\mathcal{I}}$ and $\tilde{\mathcal{K}}$.


For loss functions independent of $\bbeta$, including composite losses, the approximate loss is denoted $\lambda^*(\hat{\bbeta},\Delta)$ and the approximate external loss given by
$L^*_{DF}(\Delta) = \mathrm{E}_{\btheta \vert D,\Delta} \left[ \ell_{C}^*(\btheta,\Delta) \right]$, where $\ell^*_{C}(\btheta,\Delta) = \mathrm{E}_{\hat{\bbeta} \vert \btheta, D,\Delta} \left[ \lambda^*(\hat{\bbeta},\Delta) \right]$ is often available in closed form.

If the fitted and designer models are identical, then $\mathcal{I}_D = \mathcal{J}_D$, and $\tilde{\bbeta} = \bbeta$. In this case, we obtain an asymptotic approximation to the internal expected loss given by $L^*_{FF}(\Delta) = \mathrm{E}_{\bbeta \vert F,\Delta} \left[ \ell_{U}^*(\bbeta,\Delta) \right]$, where $\ell_{U}^*(\bbeta,\Delta)$ is the expected value of $\lambda^*(\bbeta, \hat{\bbeta},\Delta)$ with respect to (\ref{eqn:berger}).

\subsection{Exemplar loss functions}

Table~\ref{tab:loss2} shows the functions $\ell^*_{G}(\btheta,\Delta)$ and $\ell^*_{C}(\btheta,\Delta)$ for the exemplar loss functions considered in this paper (see Table~\ref{tab:loss}). Derivation of these expressions is given in Section~SM3 of the Supplementary Material. Also shown is the function $\ell^*_{U}(\bbeta,\Delta)$ for the asymptotic approximation to the internal expected loss. Note that designs that minimise the fitted prior expectation of $\ell^*_{U}(\bbeta,\Delta)$ under self-information and squared error are commonly referred to as pseudo-Bayesian D- and A-optimal, respectively. In Table~\ref{tab:loss2}, $\tilde{\mathcal{T}}_{\theta\theta} = \left(\tilde{\mathcal{I}}_{\theta\theta} - \tilde{\mathcal{I}}_{\theta\gamma}\tilde{\mathcal{I}}_{\gamma\gamma}^{-1}\tilde{\mathcal{I}}_{\gamma\theta}\right)^{-1}$ is the approximate marginal fitted posterior variance matrix of $\btheta$, under (\ref{eqn:berger}). Additionally, $\tilde{\mathcal{S}}_{\theta\theta} = I_{p_\theta} + \tilde{\mathcal{I}}_{\theta\gamma} \tilde{\mathcal{I}}_{\gamma\gamma}^{-1}\tilde{\mathcal{I}}_{\gamma\gamma}^{-1}\tilde{\mathcal{I}}_{\gamma\theta}$. 

\begin{table}
\begin{center}
\caption{Exemplar approximate external expected loss functions. \label{tab:loss2}}
\begin{tabular}{lll} \hline \hline
 & Self-information & Squared error \\ \hline 
\\
$\ell_{G}^*(\btheta, \Delta)$ & $N_{SI}- \frac{1}{2}\log \vert \tilde{\mathcal{I}} \vert + \frac{1}{2} \Vert \btheta - \tilde{\btheta} \Vert_{2,\tilde{\mathcal{T}}^{-1}_{\theta\theta}}^2$ & 
$\mathrm{tr}\left(\tilde{\mathcal{I}}^{-1}\right)  + \Vert \btheta - \tilde{\btheta} \Vert^2_{2,\tilde{\mathcal{S}}_{\theta\theta}} $
\\
& $ + \frac{1}{2} \mathrm{tr}\left( \tilde{\mathcal{T}}^{-1}_{\theta\theta}\tilde{\mathcal{K}}_{\theta\theta}\right) + p_{\gamma}/2$  & $+ \mathrm{tr}\left[\tilde{\mathcal{S}}_{\theta \theta} \left(\tilde{\mathcal{K}}_{\theta \theta} - \tilde{\mathcal{T}}_{\theta \theta} \right)\right]$\\ 
\\ \hline \hline
& Entropy & Trace variance \\ \hline
\\
$\ell_{C}^*(\btheta,\Delta)$ & $N_{SI}- \frac{1}{2}\log \vert \tilde{\mathcal{I}} \vert + \frac{p}{2}$ 
&
$\mathrm{tr}\left(\tilde{\mathcal{I}}^{-1}\right)$\\ 
\\ \hline \hline
& D-optimality & A-optimality \\ \hline
\\
$\ell_{U}^*(\bbeta,\Delta)$ & $N_{SI}- \frac{1}{2}\log \vert \mathcal{I} \vert+ \frac{p}{2}
$ & 
$\mathrm{tr} \left( \mathcal{I}^{-1} \right)$ \\ 
\\ \hline \hline
\end{tabular}
\end{center}
\end{table}

Note that $\ell^*_{G}$, $\ell^*_{C}$ and $\ell_{U}^*$ all feature $-\log \vert \mathcal{I} \vert$ (under self-information/entropy) or $\mathrm{tr} \left( \mathcal{I}^{-1}\right)$ (under squared error/trace variance). Minimising these terms has the effect of (approximately) maximising posterior precision. For the pseudo-Bayesian objective functions these are averaged with respect to the fitted prior. Whereas for external expected loss, they are averaged with respect to the designer prior of $\btheta$ through $\tilde{\bbeta}$; a function of $\btheta$ and the best possible estimate of $\bbeta$ under the fitted model. The expressions for $\ell_{C}^*$ (i.e. under composite loss) give the entropy and trace of variance under the asymptotic distribution of $\bbeta$ under a misspecified model (\ref{eqn:berger}).

The expressions for $\ell^*_{G}(\btheta, \Delta)$ under the generator squared error and self-information loss functions feature two extra terms related to the quality of $\hat{\btheta}$ as a point estimate of the common parameters $\btheta$.  The first of these, $\Vert \btheta - \tilde{\btheta} \Vert_{2,A}^2$ where $A$ is either $\tilde{\mathcal{T}}_{\theta\theta}^{-1}$ or $\tilde{\mathcal{S}}_{\theta\theta}$, relates to the bias. This term penalises designs that lead to $\hat{\btheta}$ being (on average under the designer model) far from $\btheta$. The second term measures the difference between the variance of $\hat{\btheta}$ under the fitted model, i.e. $\tilde{\mathcal{T}}_{\theta\theta}$, and under the designer model, i.e. $\tilde{\mathcal{K}}_{\theta\theta}$. 

\section{Examples}\label{sec:examples}

Two examples are considered in this section where designs are found under our proposed design framework. In both cases, the fitted model is simpler than the designer model.  As will be seen, the designs found provide robustness through replicated design points within a linear regression setting, and also through providing space filling properties for estimating parameters under a non-linear model. In Section~SM6 in the Supplementary Material, a further example is provided where the fitted model is more complex (as measured by number of parameters) than the designer model.

\subsection{Normal linear regression vs full-treatment model} \label{sec:unit}

In this example we explore the implications of assuming a full-treatment designer model within a fitted linear regression setting. Under the fitted model, it is assumed
$$\by \vert \bgamma, \sigma^2, F,\Delta \sim \mathrm{N}\left(X \bgamma, \sigma^2 I_n\right),$$
where $X$ is an $n \times p$ model matrix (a function of $\Delta$), and $\sigma^2>0$ is a nuisance parameter. We assume a conjugate normal-inverse-gamma fitted prior distribution for $\bgamma$ and $\sigma^2$ such that $\bgamma | \sigma^2, F \sim \mathrm{N}\left(\bmu_F, \sigma^2 V_F\right)$ and $\sigma^2 | F \sim \mathrm{IG}\left(a_F/2, b_F/2\right)$. For the designer model, at this stage, it is only assumed that $\by$ has finite mean $\mathrm{E}\left(\by \vert D,\Delta\right) = \bm_D$ and variance $\mathrm{var}\left(\by \vert D,\Delta\right) = \Sigma_D$. Note that the fitted and designer models are disjoint, sharing no common parameters.

Under the fitted and designer models, it is possible to derive closed form expressions, to some extent, for the external expected loss. The fitted posterior distribution of $\bgamma$ is $\mathrm{t}\left(\hat{\bmu}_F,  \hat{b}_F \hat{V}_F/\hat{a}_F, \hat{a}_F\right)$, i.e. a multivariate t-distribution with mean $\hat{\bmu}_F = \hat{V}_F \left(X^\T  \by + V_F^{-1} \bmu_F\right)$, scale matrix $\hat{b}_F\hat{V}_F/\hat{a}_F$, and $\hat{a}_F = a_F+n$ degrees of freedom, where 
$$
\hat{V}_F = \left(X^\T X + V_F^{-1} \right)^{-1}; \qquad \hat{b}_F = b_F + \Vert \by - X \bmu_F \Vert^2_{2,\Sigma_F^{-1}}; \qquad \Sigma_F = I_n + XV_FX^\T .
$$
The external expected self-information and squared error loss functions are given by
\begin{eqnarray}
L_{DF,SI}(\Delta) & = & H_{SI,1} + \frac{1}{2} \log \vert \hat{V}_F\vert + \frac{p}{2} \mathrm{E}_{\by\vert D,\Delta}\left(\log\hat{b}_F\right),\label{eqn:lmsi} \\
L_{DF,SE}(\Delta) & = & \frac{\mathrm{E}_{\by\vert D,\Delta}\left(\hat{b}_F\right)}{\hat{a}_F-2} \mathrm{tr}\left(\hat{V}_F\right),\label{eqn:lmse}
\end{eqnarray}
respectively, where $\mathrm{E}_{\by\vert D,\Delta}(\hat{b}_F) = b_F + \Vert \bm_D - X \bmu_F \Vert^2_{2,\Sigma_F^{-1}} + \mathrm{tr}\left( \Sigma_D \Sigma_F^{-1}\right)$ and $H_{SI,1}$ is a constant not depending on the design. Derivations of (\ref{eqn:lmsi}) and (\ref{eqn:lmse}) are provided in Section~SM4 in the Supplementary Material. The expectation of $\log\hat{b}_F$ with respect to the designer model, appearing in the external expected self-information loss (\ref{eqn:lmsi}), is not available in closed form. In what follows, we use a delta approximation \citep[for example][pages 33 -35]{davison2003} where $\mathrm{E}_{\by\vert D,\Delta}(\log\hat{b}_F) \approx \log \mathrm{E}_{\by\vert D,\Delta}(\hat{b}_F)$. However, note that by Jensen's inequality, i.e. $\mathrm{E}_{\by\vert D,\Delta}(\log\hat{b}_F) \le \log \mathrm{E}_{\by\vert D,\Delta}(\hat{b}_F)$, the approximate external expected self-information loss (which we minimise to find designs) is actually an upper bound on the true external expected loss.

Compare (\ref{eqn:lmsi}) and (\ref{eqn:lmse}) to the corresponding expressions for the internal expected self-information and squared error given by
$$
L_{FF,SI}(\Delta) = H_{SI,2} + \frac{1}{2} \log \vert \hat{V}_F\vert , \qquad \mbox{and} \qquad
L_{FF,SE}(\Delta) = \frac{b_F}{a_F-2} \mathrm{tr}\left(\hat{V}_F\right),
$$
respectively, where $H_{SI,2}$ is a constant not depending on the design. The difference is that the external expected loss functions feature expectation of (a function of) $\hat{b}_F$ under the designer model. \citet[][page 319]{ohaganforster2004} describe how $\hat{b}_F$ summarises the inadequacy of the fitted model so it is natural that this drives the difference, with the external expected losses favouring designs for which the expected value of $\hat{b}_F$ is small.

Now assume a specific form for the designer model as a full-treatment model \citep[e.g][]{gilmourtrinca2012}, i.e. where each unique treatment permits a unique mean response. Mathematically, the full-treatment designer model is 
$$
\by \vert \btau, \sigma^2, D,\Delta \sim \mathrm{N}\left( Z \btau, \sigma^2 I_n \right); \quad \btau \vert \sigma^2, D,\Delta \sim  \mathrm{N}\left( \bmu_D, \sigma^2 V_D \right); \quad  \sigma^2 \vert D \sim  \mathrm{IG}\left(\frac{a_D}{2},\frac{b_D}{2}\right),
$$
where $\btau$ is a $q \times 1$ vector of mean treatment effects. The $n \times q$ designer model matrix $Z$ is a function of $\Delta$ where $q$ is the number of unique treatments. For $i=1,\dots,n$, the $i$th row of $Z$ is the vector $\be_j$, i.e. a vector of zeros except for the $j$th element which is one, indicating that run $i$ receives unique treatment $j$. Therefore, $\bm_D = Z\bmu_D$, $\Sigma_D = b_D \left(I_n + ZV_DZ^\T \right)/(a_D-2)$ and $\mathrm{E}_{\by\vert D,\Delta}(\hat{b}_F) = b_F + \Vert Z\bmu_D - X \bmu_F \Vert^2_{2,\Sigma_F^{-1}} + \mathrm{tr}\left( \Sigma_D \Sigma_F^{-1}\right)$.

Interpretation is simplified by assuming non-informative fitted and designer prior distributions where, for the fitted model $b_F=0$, $\bmu_F = \bzero_p$ (the $p \times 1$ vector of zeros) and $V_F^{-1} = 0_{p \times p}$ (the $p \times p$ matrix of zeros), and, for the designer model, $\bmu_D=\bzero_q$. However, $V_D$ needs to be finite for $\Sigma_D$ to be finite and for the external expected loss to exist. It is assumed the elements of $\btau$ are independent with $\tau_j \sim \mathrm{N}\left(0, \kappa \sigma^2/n_j \right)$, where $\kappa>0$ and $n_j$ is the number of runs receiving the $j$th unique treatment, for $j=1,\dots,q$. Under this prior specification, the designer posterior mean of $\tau_j$ is the sample mean of the corresponding responses shrunk toward zero by a common factor $\kappa/(1+\kappa)$, with $\kappa$ controlling the amount of shrinkage. Consequently, $V_D = \kappa \left(Z^\T  Z\right)^{-1}$.

Under these prior specifications, minimising the external expected self-information and squared error loss functions is equivalent to minimising
\begin{eqnarray}
\bar{L}_{DF,SI}(\Delta) & = & p \log \left[(1+\kappa)(n - p) - \kappa d\right] - \log \vert X^\T X \vert, \label{eqn:lmsi3} \\
\bar{L}_{DF,SE}(\Delta) & = & \left[(1+\kappa)(n - p) - \kappa d\right)] \mathrm{tr}\left[ \left(X^\T X\right)^{-1}\right] \label{eqn:lmse3}
\end{eqnarray}
respectively, where $d=n-q$ is known as the pure error degrees of freedom, i.e. the number of repeated treatments. Derivations of expressions (\ref{eqn:lmsi3}) and (\ref{eqn:lmse3}) are provided in Section~SM5 of in the Supplementary Material. We refer to designs that minimise (\ref{eqn:lmsi3}) and (\ref{eqn:lmse3}) as DE- and AE-optimal, respectively. The corresponding expressions to (\ref{eqn:lmsi3}) and (\ref{eqn:lmse3})  under the internal expected loss are
$$\bar{L}_{FF,SI}(\Delta) =- \log \vert X^\T X \vert, \qquad \mbox{and} \qquad \bar{L}_{FF,SE}(\Delta) = \mathrm{tr}\left[ \left(X^\T X\right)^{-1}\right],$$
respectively, and designs that minimise $\bar{L}_{FF,SI}(\Delta)$ and $\bar{L}_{FF,SE}(\Delta)$ are called D- and A-optimal, respectively. Hence it can be seen that under the external expected loss there is a trade-off between precision of estimation of $\bgamma$, as measured by functions of $X^\T X$, and pure error degrees of freedom, i.e. DE- and AE-optimal designs will tend to feature more replication than D- and A-optimal designs, respectively. A numerical example verifying this is now provided.

Consider Example 1 from \citet{gilmourtrinca2012} involving an experiment with $n=16$ runs and $k=3$ design variables. The fitted model is a second-order model including an intercept, three first-order terms, three quadratic terms and three pairwise interactions, i.e. $p = 10$. We specify $\kappa = n$ leading to a unit-information prior for $\btau$ \citep{smithspiegelhalter1980}.

DE- and AE-efficiency of a design $\Delta$ are defined by
\begin{eqnarray*}
\mathrm{Eff}_{DE}(\Delta) &=& \exp\left\{ \frac{1}{p} \left[\bar{L}_{DF,SI}(\Delta^*_{DF,SI}) - \bar{L}_{DF,SI}(\Delta)\right] \right\}\\
\mathrm{Eff}_{AE}(\Delta) &=& \frac{\bar{L}_{DF,SE}(\Delta^*_{DF,SE})}{\bar{L}_{DF,SE}(\Delta)},
\end{eqnarray*}
respectively, where $\Delta^*_{DF,SI}$ and $\Delta^*_{DF,SE}$ are the DE- and AE-optimal designs, respectively, with similar expressions for D- and A-efficiency. The D-efficiency of the DE-optimal design is 85\%, whereas the DE-efficiency of the D-optimal design is only 7\%. The corresponding figures for squared error are 84\% and 10\%. Clearly, the D- and A-optimal designs are not robust to the external expected loss, whereas the DE- and AE-optimal designs are robust to the internal expected loss. The D-optimal design has $q=16$ unique treatments compared to $q=10$ for the DE-optimal design. The equivalent values for the A- and AE-optimal designs are $q=14$ and $q=10$, respectively. The minimum number of unique treatments to be able to estimate all elements of $\bgamma$ under the fitted model is $q=p=10$. Thus the DE- and AE-optimal designs feature the maximum amount of replication possible whilst still being able to estimate $\bgamma$.

Expressions for the external expected loss under self-information (\ref{eqn:lmsi3}) and squared error (\ref{eqn:lmse3}) should be compared to the following objective functions under the DP- and AP-criteria \citep{gilmourtrinca2012} 
\begin{eqnarray}
\bar{L}_{DP}(\Delta) & = & p\log F_{p,d,1-\alpha}- \log \vert X^\T X \vert  \label{eqn:DP} \\
\bar{L}_{AP}(\Delta) & = & F_{1,d,1-\alpha}\mathrm{tr}\left[ \left(X^\T X\right)^{-1}\right]. \label{eqn:AP}
\end{eqnarray}
respectively. Note that $\bar{L}_{DP}(\Delta)$ in (\ref{eqn:DP}) is actually the natural logarithm of the objective function in \citet{gilmourtrinca2012} but the optimisation problem is equivalent due to $\log$ being a monotonically increasing function. In (\ref{eqn:DP}) and (\ref{eqn:AP}), $F_{p,d,1-\alpha}$ the $1-\alpha$-quantile of the $F_{p,d}$ distribution with $F_{p,d,1-\alpha}$ being a decreasing function of $d$. The motivation behind the DP- and AP-criteria is to minimise the volume of the confidence region or sum of confidence intervals, respectively, for $\bgamma$, where the response variance $\sigma^2$ is estimated under the full-treatment model; a model-independent estimate. In both cases, the objective functions are modified by a decreasing function of the pure error degrees of freedom, $d$. As shown above we are able to provide a different, Bayesian decision-theoretic, motivation for the same outcome.

\subsection{Estimating parameters of a Michaelis-Menten model under model discrepancy} \label{sec:nlm}

In this section we consider designing an experiment to estimate the parameters of a non-linear model. Finding designs for such models is widely featured in the design of experiments literature (see, e.g. \citealt{pronzato_pazman_2013}; \citealt{federov_leonov_2014}). Under such models, it is hypothesised that the expectation of $y_i$ is $\eta(\btheta, \Delta_i)$; a mathematical model based on mechanistic or phenomenological arguments. As such the parameters $\btheta$ can have interpretable physical meaning. 

As a specific example, consider the Michaelis-Menten model \citep[e.g.][]{dette_biedermann_2003}, commonly-used in the study of pharmacokinetics and chemical kinetics. A standard experiment has a response given by reaction velocity $y_i$ measured for $k=1$ controllable variable: the scaled concentration of substrate $\Delta_i = x_i \in [0,1]$, for $i=1,\dots,n$. The model has 
\begin{equation}
\eta(\btheta, x) = \theta_1 xL/\left(xL+\theta_2\right),
\label{eqn:mmmodel}
\end{equation}
where $\theta_1>0$ is the maximum velocity of the chemical reaction, $\theta_2>0$ is the substrate concentration at which the reaction rate is half of $\theta_1$, and $L$ is the maximum possible concentration of substrate, which we assume is $L=400$.

Supposing a normal distribution, we assume the following fitted model
\begin{equation}
\by \vert \btheta, \sigma^2, F, \Delta \sim \mathrm{N}\left( \bseta(\btheta, \Delta), \sigma^2 I_n \right)
\label{eqn:nlmfitted}
\end{equation}
where $\bseta(\btheta, \Delta) = \left(\eta(\btheta, x_1), \dots, \eta(\btheta, x_n)\right)^\T$ and $\sigma^2 > 0$ is the unknown response variance (a nuisance parameter, not of direct interest). The design problem is to choose the values of $\Delta = \left(x_1,\dots,x_n\right)^\T$, i.e. the $n$ concentrations.

The fitted model given by (\ref{eqn:nlmfitted}) is assumed for pragmatic convenience. However the reasoning behind the mathematical model is usually incomplete, meaning that there can be a systematic discrepancy between the true expected value of $y_i$ and $\eta(\btheta, x_i)$. The \cite{kennedyohagan2001} framework aims to model this discrepancy using a Gaussian process and we assume this for the designer model. Specifically, the designer model is assumed to be 
\begin{equation}
\by \vert \btheta, \sigma^2, \bdelta, D, \Delta \sim \mathrm{N}\left( \bseta(\btheta, \Delta) + \bdelta, \sigma^2 I_n \right).
\label{eqn:nlmdesigner}
\end{equation}
where 
\begin{equation}
\bdelta \vert \alpha, \rho, \sigma^2, D, \Delta \sim \mathrm{N} \left( \bzero_n, \sigma^2 \rho C \right).
\label{eqn:nlmdesignerdisc}
\end{equation}
In (\ref{eqn:nlmdesignerdisc}), $\rho>0$, $\tau^2>0$, and $C$ is an $n \times n$ correlation matrix with $ij$th element $C_{ij} = c(\vert x_i - x_j \vert ; \alpha)$ where $c( \vert x_i- x_j \vert ; \alpha)$ is a correlation function depending on $\alpha>0$ and the distance between $x_i$ and $x_j$, with the property that $c( 0 ; \alpha) = 1$. In what follows, we choose the M{\'a}tern $\left(\frac{5}{2}\right)$ correlation function where
$$c(d ; \alpha) = \left( 1 + \frac{d}{\alpha} + \frac{d^2}{3 \alpha^2} \right) \exp \left( - \frac{d}{\alpha} \right)$$
a commonly-used correlation function \citep[e.g.][page 84]{rasmussen_williams_2006}. It follows from (\ref{eqn:nlmdesigner}) and (\ref{eqn:nlmdesignerdisc}) that
\begin{equation}
\by \vert \btheta, \sigma^2, \alpha, \rho, D, \Delta \sim \mathrm{N}\left[\bseta(\btheta, \Delta) , \sigma^2 \left( I_n + \rho C \right) \right].
\label{eqn:nlmdesigner2}
\end{equation}

We consider three designs found by minimising
$$
\begin{array}{llll}
\mbox{(i)} & \mbox{external expected squared error loss;} & \mbox{(ii)} & \mbox{external expected trace variance loss;}\\
\mbox{(iii)} & \mbox{internal expected squared error loss.} & &
\end{array}
$$
In each case, the expected loss is not available in closed form. Indeed, since the fitted posterior is not of known form, the loss function itself is analytically intractable. To approach this problem, we use a nested Monte Carlo approximation \citep[e.g.][]{rainforth2018}. The following algorithm is used to approximate the external expected squared error loss.
\begin{enumerate}
\item
Generate a sample $\left\{ \btheta_b, \sigma^2_b, \alpha_b, \rho_b \right\}_{b=1}^B$ of size $B$ from the designer prior distribution of $\btheta$, $\sigma^2$, $\alpha$ and $\rho$.
\item
For $b=1,\dots,B$, generate $\by_b$ from the designer likelihood 
$$\by \vert \btheta_b, \sigma_b^2, \alpha_b, \rho_b, D, \Delta \sim \mathrm{N}\left[ \bseta(\btheta_b, \Delta) , \sigma_b^2 \left( I_n + \rho_b C_b \right) \right].$$
\item \label{step3}
Approximate the external expected squared error loss by the following Monte Carlo approximation
\begin{equation}
\hat{L}_{DF,SE}(\Delta) = \frac{1}{B} \sum_{b=1}^B \Vert \btheta_b - \hat{\mathrm{E}} \left(\btheta \vert \by_b, F, \Delta \right) \Vert^2_{2, I_2}.
\label{eqn:dlmc}
\end{equation}
In (\ref{eqn:dlmc}), $\hat{\mathrm{E}} \left(\btheta \vert \by_b, F, \Delta \right)$ is a Monte Carlo approximation to the fitted posterior mean of $\btheta$ conditional on $\by_b$. This approximation is formed as follows. Generate a sample $\left\{ \tilde{\btheta}_j, \tilde{\sigma}_j^2 \right\}_{j=1}^{\tilde{B}}$ of size $\tilde{B}$ from the fitted prior distribution of $\btheta$ and $\sigma^2$. Then 
$$\hat{\mathrm{E}} \left(\btheta \vert \by_b, F, \Delta \right) = \frac{\sum_{j=1}^{\tilde{B}} \tilde{\btheta}_j \pi(\by_b \vert \tilde{\btheta}_j, \tilde{\sigma}_j^2, F, \Delta)}{\sum_{j=1}^{\tilde{B}} \pi(\by_b \vert \tilde{\btheta}_j, \tilde{\sigma}_j^2, F, \Delta)},$$
for $b=1,\dots,B$.
\end{enumerate}
Note that the sample generated in Step \ref{step3} is reused to calculate $\hat{\mathrm{E}} \left(\btheta \vert \by_b, F, \Delta \right)$ for all $b=1,\dots,B$.

In a similar fashion, the external expected trace variance loss can be approximated by
$$\hat{L}_{DF,TV}(\Delta) = \frac{1}{B} \sum_{b=1}^B \mathrm{tr} \left\{ \hat{\mathrm{var}} \left( \btheta \vert \by_b, F, \Delta \right) \right\}$$
with
$$\hat{\mathrm{var}} \left( \btheta \vert \by_b, F, \Delta \right) = \hat{\mathrm{E}} \left(\btheta \btheta^T \vert \by_b, F, \Delta \right) - \hat{\mathrm{E}} \left(\btheta \vert \by_b, F, \Delta \right)\hat{\mathrm{E}} \left(\btheta \vert \by_b, F, \Delta \right)^{\T},$$
and
$$\hat{\mathrm{E}} \left(\btheta \btheta^{\T} \vert \by_b, F, \Delta \right) = \frac{\sum_{j=1}^{\tilde{B}} \tilde{\btheta}_j \tilde{\btheta}_j^{\T} \pi(\by_b \vert \tilde{\btheta}_j, \tilde{\sigma}_j^2, F, \Delta)}{\sum_{j=1}^{\tilde{B}} \pi(\by_b \vert \tilde{\btheta}_j, \tilde{\sigma}_j^2, F, \Delta)}.$$

The internal expected squared error loss is approximated by, first, generating a sample $\left\{ \btheta_b, \sigma_b^2 \right\}_{b=1}^B$ of size $B$ from the fitted prior distribution of $\btheta$ and $\sigma^2$. For $b=1, \dots, B$, generate $\by_b$ from the fitted likelihood
$$\by \vert \btheta_b, \sigma_b^2, F, \Delta \sim \mathrm{N}\left( \bseta(\btheta_b, \Delta), \sigma_b^2 I_n \right).$$
Then
$$\hat{L}_{FF,SE}(\Delta) = \frac{1}{B} \sum_{b=1}^B \Vert \btheta_b - \hat{\mathrm{E}} \left(\btheta \vert \by_b, F, \Delta \right) \Vert^2_{2, I_2}.$$

We find designs with $n=20$ concentrations. The fitted prior distribution is such that $\btheta$ and $\sigma^2$ are independent with $\theta_1, \theta_2 \stackrel{\mathrm{iid}}{\sim} \mathrm{U}[20,200]$ and $\sigma^2 \sim \mathrm{Exp}(1)$; an exponential distribution with mean one. The designer prior distribution is such that $\btheta$, $\sigma^2$, $\alpha$ and $\rho$ are independent with $\theta_1, \theta_2 \stackrel{\mathrm{iid}}{\sim} \mathrm{U}[20,200]$, $\sigma^2, \rho \stackrel{\mathrm{iid}}{\sim} \mathrm{Exp}(1)$ and $\alpha \sim \mathrm{Exp}(5)$.

\begin{table}[t]
\begin{center}
\caption{Efficiencies and number of unique concentrations for the three designs found for the Michaelis-Menten model. \label{tab:mm}}

\vspace{1mm}

\begin{tabular}{lrrrr} \hline \hline
Design & \multicolumn{3}{c}{Efficiency} & \multicolumn{1}{l}{Number of unique} \\
 & 	\multicolumn{3}{c}{}	& \multicolumn{1}{l}{concentrations} \\ \hline
& \multicolumn{1}{c}{External} & \multicolumn{1}{c}{External} & \multicolumn{1}{c}{Internal} & \\
& \multicolumn{1}{c}{SE (i)} & \multicolumn{1}{c}{TV (ii)} & \multicolumn{1}{c}{SE (iii)} & \\ \hline \hline
External SE (i) & 100 & 67.0 & 79.7 & 13 \\
External TV (ii) & 95.4 & 100.0 & 99.9 & 12 \\
Internal SE (iii) & 86.1 & 85.8 & 100 & 12 \\ \hline \hline
\end{tabular}
\end{center}

\end{table}

To minimise the nested Monte Carlo approximations to the expected loss functions, we use the approximate coordinate exchange algorithm \citep{overstall_woods_2017}. We use $B = \tilde{B} = 20,000$ but it is noted that if $\Vert \btheta - \hat{\mathrm{E}} \left(\btheta \vert \by, F, \Delta \right) \Vert^2_{2, I_2}$ is continuously differentiable then the convergence rate can be slightly improved \citep{rainforth2018} by setting $B \propto \tilde{B}^2$ (whilst holding $B\tilde{B}$ fixed).

\begin{center}
\begin{figure}
\caption{Designs for the Michaelis-Menten model. The top panel shows $\mu(\btheta, x)$ plotted against $x$ for 100 values generated from the designer prior of $\btheta$. The bottom panel shows the location of the concentrations for each of the three designs, i.e. found by minimising (i) external expected squared error loss; (ii) external expected trace variance loss; and (iii) internal expected squared error loss. The size of the plotting symbol is an increasing function of the replication of each concentration.} \label{fig:mm} 
\includegraphics[scale=0.475]{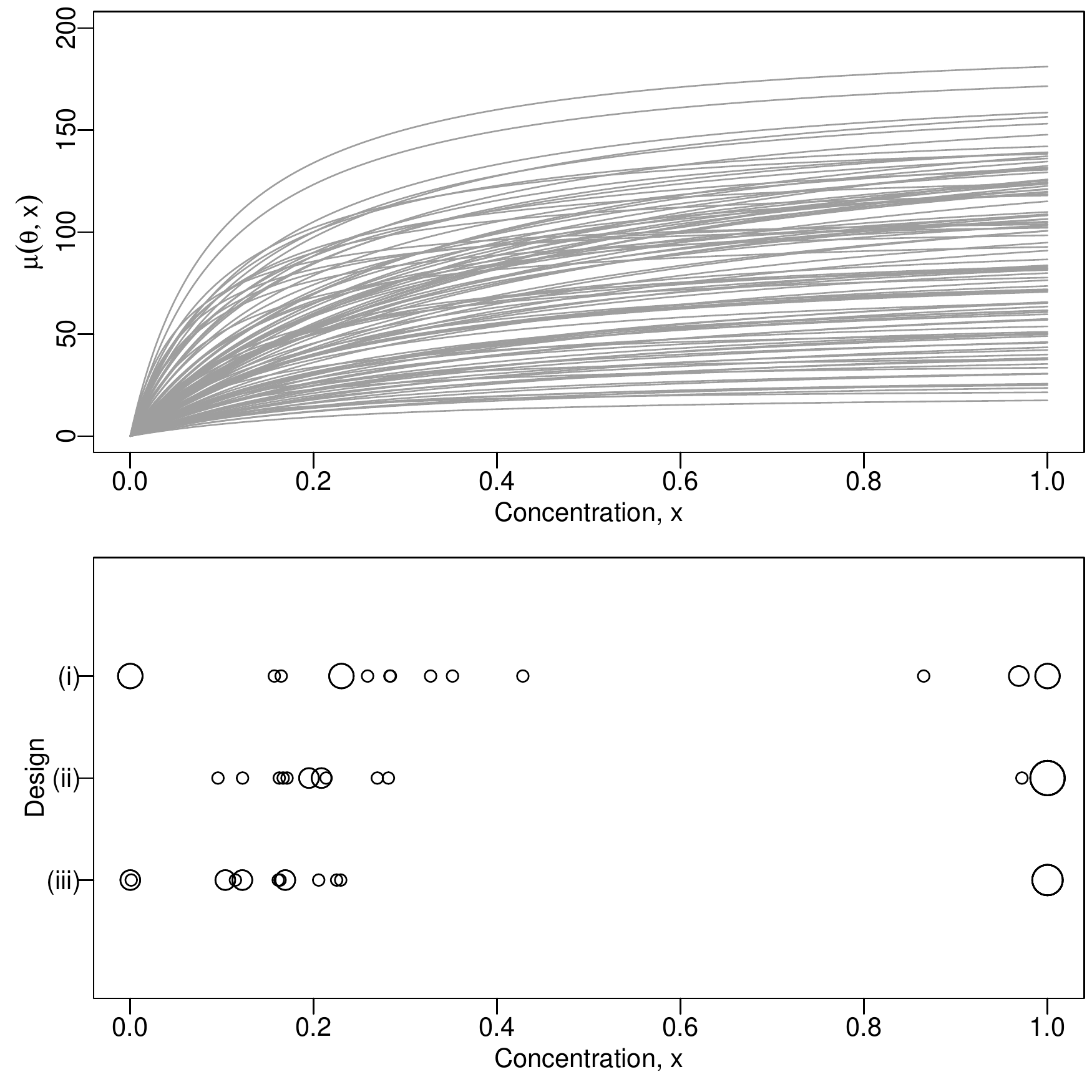}
\end{figure}
\end{center}

\vspace{-10mm}

Table~\ref{tab:mm} shows the efficiencies of the three designs under the three expected loss functions and the number of unique concentrations. The top panel of Figure~\ref{fig:mm} shows $\mu(\btheta, x)$ plotted against $x$ for 100 values generated from the designer prior of $\btheta$ and the bottom panel shows the location of the concentrations for each of the three designs, i.e. found by minimising (i) external expected squared error loss; (ii) external expected trace variance loss; and (iii) internal expected squared error loss. The size of the plotting symbol is an increasing function of the replication of each concentration. From Figure~\ref{fig:mm}, the three designs have common clusters of concentrations around 0.2 and 1. The designs found by minimising the external expected trace variance and internal expected squared error loss functions are broadly similar. This is confirmed in Table~\ref{tab:mm} by the relatively high efficiencies for these designs. Compared to the other designs, the concentrations under the design that minimises the external expected squared error loss are more ``spread out", i.e. a compromise between the designs found under external trace variance and internal squared error and a space-filling design.
 
The similarity of the designs found under the external expected trace variance (ii) and the internal expected squared error (iii) loss functions agrees with the asymptotic results from Section~\ref{sec:nlm}. In this example, the function $d(\bt)$ given by (\ref{eqn:dfunc}) is
\begin{eqnarray*}
d(\bt_\theta, t_{\sigma^2}) & = & \mathrm{E}_{\sigma^2, \rho, \alpha \vert D, \Delta} \left\{ \mathrm{E}_{\by \vert \btheta, \sigma^2, \rho, \alpha, D, \Delta} \left[ \log \pi(\by \vert \bt_{\theta}, t_{\sigma^2}, F, \Delta) \right] \right\}\\
& = & - \frac{n}{2} \log (2\pi) - \frac{n}{2} \log t_{\sigma^2} - \frac{1}{2 t_{\sigma^2}} \Vert \bseta (\btheta, \Delta) - \bseta (\bt_{\theta}, \Delta) \Vert^2_{2,I_n} - \frac{n \sigma_0^2 \left( 1 + \rho_0 \right)}{2 t_{\sigma^2}}
\end{eqnarray*}
where $\bt_\theta$ and $t_{\sigma^2}$ are the elements of $\bt$ corresponding to $\btheta$ and $\sigma^2$, respectively, and $\sigma^2_0 = \mathrm{E}_{\sigma^2 | D, \Delta} (\sigma^2)$ and $\rho_0 = \mathrm{E}_{\rho | D, \Delta} (\rho)$ are the designer prior means of $\sigma^2$ and $\rho$, respectively. Clearly, $d(\bt_\theta, t_{\sigma^2})$ is maximised by $\tilde{\btheta} = \btheta$ and $\tilde{\sigma}^2 = \sigma_0^2 \left( 1 + \rho_0 \right)$. It then follows from (\ref{eqn:berger}) that the fitted marginal posterior distribution of $\btheta$ is approximated by 
$$\mathrm{N} \left(\btheta, \tilde{\mathcal{I}}_{\theta \theta}^{-1} \right),$$
where
$$\tilde{\mathcal{I}}_{\theta \theta}^{-1} = \sigma_0^2\left( 1 + \rho_0 \right) \left[ \sum_{i=1}^n \left. \frac{\partial \eta(\bt, \Delta_i)}{\partial \bt} \right\vert_{\bt = \btheta} \left. \frac{\partial \eta(\bt, \Delta_i)}{\partial \bt^\T} \right\vert_{\bt = \btheta} \right]^{-1}.$$
Therefore, from Table~\ref{tab:loss2}, the external expected trace variance loss is approximated by 
\begin{equation}
L^*_{DF,TV}(\Delta) = \mathrm{E}_{\btheta \vert D, \Delta} \left[ \mathrm{tr} \left( \tilde{\mathcal{I}}_{\theta \theta}^{-1} \right) \right].
\label{eqn:eq1}
\end{equation}
By contrast, from Table~\ref{tab:loss2}, the internal expected squared error loss is approximated by
\begin{equation}
L^*_{FF,SE}(\Delta) = \mathrm{E}_{\btheta \vert F, \Delta} \left[ \mathrm{tr} \left( \mathcal{I}_{\theta \theta}^{-1} \right) \right],
\label{eqn:eq2}
\end{equation}
where
$$\mathcal{I}_{\theta \theta}^{-1} = \mathrm{E}_{\sigma^2 | F, \Delta} (\sigma^2)  \left[ \sum_{i=1}^n \left. \frac{\partial \eta(\bt, \Delta_i)}{\partial \bt} \right\vert_{\bt = \btheta} \left. \frac{\partial \eta(\bt, \Delta_i)}{\partial \bt^\T} \right\vert_{\bt = \btheta} \right]^{-1}.$$
Since $\sigma_0^2\left( 1 + \rho_0 \right)$ and $\mathrm{E}_{\sigma^2 | F, \Delta}(\sigma^2)$ do not depend on the design (for this example), the approximations (\ref{eqn:eq1}) and (\ref{eqn:eq2}) are minimised by the same design.

\section{Discussion}

An extended framework is proposed for Bayesian design whereby the expectation of the loss is taken with respect to the probability distribution implied by an alternative model (the designer model) to the one being fitted to the observed responses. Properties of the framework are explored. The key findings are that the external expected loss is an upper bound on the corresponding internal expected loss under the designer model, providing justification for using a simpler fitted model for the purposes of computing a design. We also observed that designs found under composite and generator loss functions are no longer equivalent. By developing an asymptotic approximation to the external expected loss we were able to show that the composite loss functions focus solely on maximising posterior precision, whereas generator loss functions also incorporate point estimation.


%

We encourage researchers consider the extended framework when designing experiments under the Bayesian decision-theoretic approach, if not to specify the design to be used in practice, but at least as a benchmark to assess the performance of that design. Designs can be found under the proposed framework with only minor modification of existing computational methods (see Section~\ref{sec:nlm} and Section~SM6 in the Supplementary Material).

In two of the examples, the designer model is more complex than the fitted model, and we are, in a sense, attempting to insert robustness into the design procedure. Robust Bayesian inference (i.e. inference under the M-open paradigm of \citealt{bernardo_smith_1994}, page 385) has become an important research field in recent years \citep[e.g.][]{bissiri_etal_2016, safebayes, jewson_etal_2018}. Related to this is a body of current work on the estimation of the parameters of a non-linear model (see Section~\ref{sec:nlm}) taking account of the incompleteness of the underlying mathematical model \citep[e.g.][]{plumlee_2017, xie_xu_2021}. Considering Bayesian decision-theoretic design of experiments under these approaches will be the focal point of future work.

Lastly, we have focused on an experimental aim of parameter estimation. In principle, the extended framework can be generalised to model selection and prediction aims. Exploring this will be a further avenue for future work.

\section*{Acknowledgement}
The authors would like to thank the associate editor and two anonymous reviewers for providing comments which led to substantial improvement in the paper. They would also like to thank members of the Design Study Group at the University of Southampton for initial discussions and feedback.  JM was supported by an Australian Research Council Discovery Project (DP200101263).

\newpage

  \renewcommand{\thesection}{SM\arabic{section}}
 \renewcommand{\thefigure}{SM\arabic{figure}}
 \renewcommand{\thetable}{SM\arabic{table}}
  \renewcommand{\theequation}{SM\arabic{equation}}

\setcounter{section}{0}
\setcounter{figure}{0}
\setcounter{table}{0}
\setcounter{equation}{0}

\section{Proof of Lemma 2.1} \label{sec:lemma21}

Recall that the fitted and designer models are compatible, so that $\bbeta = \btheta$. Therefore, the internal expected loss under the designer model can be written as
\begin{eqnarray}
L_{DD}(\Delta) & = & \mathrm{E}_{\by, \bbeta \vert D,\Delta} \left[ \lambda(\bbeta, \by, D,\Delta) \right] \nonumber \\
& = &\mathrm{E}_{\by \vert D,\Delta} \left\{ \mathrm{E}_{\bbeta \vert \by, D,\Delta} \left[ \lambda(\bbeta, \by, D,\Delta)\right]\right\}\nonumber \\
& = & \mathrm{E}_{\by \vert D,\Delta} \left\{ \mathrm{E}_{\bbeta \vert \by, D,\Delta} \left[ \lambda_0(\bbeta, \hat{\bb}_D)\right]\right\}, \label{eqn:cr}
\end{eqnarray}
where $\hat{\bb}_D = \arg \min_{\bb \in \mathcal{B}} \mathrm{E}_{\bbeta \vert \by, D,\Delta} \left[ \lambda_0(\bbeta, \bb)\right]$. Since $\bb = \hat{\bb}_D$ minimises $\mathrm{E}_{\bbeta \vert \by, D,\Delta} \left[ \lambda_0(\bbeta, \bb)\right]$ over $\mathcal{B}$, it follows that
$$\mathrm{E}_{\bbeta \vert \by, D,\Delta} \left[ \lambda_0(\bbeta, \hat{\bb}_D)\right] \le \mathrm{E}_{\bbeta \vert \by, D,\Delta} \left[ \lambda_0(\bbeta, \hat{\bb}_F)\right],$$
where $\hat{\bb}_F = \arg \min_{\bb \in \mathcal{B}} \mathrm{E}_{\bbeta \vert \by, F,\Delta} \left[ \lambda_0(\bbeta, \bb)\right]$. Since $\lambda(\bbeta, \by, F,\Delta) = \lambda_0(\bbeta, \hat{\bb}_F)$, it follows that
$$\mathrm{E}_{\bbeta \vert \by, D,\Delta} \left[ \lambda_0(\bbeta, \hat{\bb}_D)\right] \le \mathrm{E}_{\bbeta \vert \by, D,\Delta} \left[ \lambda(\bbeta, \by, F,\Delta)\right],$$
and from (\ref{eqn:cr}), that
$$
L_{DD}(\Delta) \le \mathrm{E}_{\by \vert D,\Delta} \left\{\mathrm{E}_{\bbeta \vert \by, D,\Delta} \left[ \lambda(\bbeta, \by, F,\Delta)\right]\right\} = L_{DF}(\Delta),$$
for compatible models.

\section{Proof of Theorem~3.1} \label{sec:theorem31}

The following regularity conditions are sufficient to prove Theorem~3.1.

\begin{enumerate}
\item[A1]
$\bbeta \in \mathcal{B}$ where $\mathcal{B}$ is an open, convex and bounded subset of $\mathbb{R}^p$.
\item[A2]
The fitted likelihood satisfies \emph{local asymptotic normality} (LAN; see, for example, \citealt{vandervaart2012}, Chapter 7).
\item[A3]
The fitted log-likelihood $\log(\by \vert \bt, F, \Delta)$ converges almost surely to $d(\bt)$ for all $\bt \in \mathcal{B}$.
\item[A4]
$\tilde{\bbeta} \in \mathcal{B}$ uniquely maximises $d(\bt)$.
\item[A5]
For $\mathcal{A}_\epsilon = \left\{ \bt \in \mathcal{B} : d(\bt) > d(\tilde{\bbeta}) - \epsilon \right\}$,
$$\int_{\mathcal{A}_\epsilon} \pi(\bbeta \vert F, \Delta) \mathrm{d}\bbeta > 0,$$
for all $\epsilon >0$.
\item[A6]
The fitted prior density $\pi(\bbeta \vert F, \Delta)$ is continuous and positive at $\tilde{\bbeta}$.
\item[A7]
$- \left. \partial^2 \log \pi (\by \vert \bt, F, \Delta)/ \partial \bt \partial \bt^\T \right\vert_{\bt = \tilde{\bbeta}}$ converges in probability to $\tilde{\mathcal{I}}$.
\item[A8]
$\left. \partial \log \pi (\by \vert \bt, F, \Delta)/ \partial \bt \right\vert_{\bt = \tilde{\bbeta}}$ converges in distribution to $\mathrm{N} \left( \bzero_p, \tilde{\mathcal{J}} \right)$.
\end{enumerate}

By assumption of conditions A3-A5, Theorem 2.2 of \citet{miller2019} establishes a Bayesian consistency for the fitted posterior distribution, i.e. 
\begin{equation}
\int_{\mathbb{A}_\epsilon} \pi(\bbeta \vert \by, F, \Delta) \mathrm{d}\bbeta \to 1
\label{eqn:consistency}
\end{equation}
as $n \to \infty$. 

Then, by assumption of conditions A1 and A2, Theorem 2.1 of \citet{kleijn_vandervaart_2012} shows that the fitted posterior distribution converges to (5) in total variation, as $n \to \infty$.

The external expected loss given by (2) is approximated by
\begin{eqnarray*}
L^*_{DF}(\Delta) &=& \mathrm{E}_{\btheta, \by \vert D,\Delta} \left\{ \mathrm{E}_{\bgamma| \by, \btheta, F,\Delta} \left[ \lambda^*(\bbeta, \hat{\bbeta}, \Delta)\right] \right\}\\
& = & \mathrm{E}_{\btheta \vert D,\Delta} \left( \mathrm{E}_{\by \vert \btheta, D,\Delta} \left\{ \mathrm{E}_{\bgamma| \by, \btheta, F,\Delta} \left[ \lambda^*(\bbeta, \hat{\bbeta}, \Delta)\right] \right\} \right),
\end{eqnarray*}
by replacing the loss by the approximate loss under (5). 

Consider the inner expectation of $\lambda^*(\bbeta, \hat{\bbeta}, \Delta)$ with respect to $\bgamma| \by, \btheta, F,\Delta$. Using properties of the multivariate normal and block matrices \citep[e.g.][page 95]{gentle2007}, it is straightforward to derive
$$\bgamma \vert \hat{\bbeta}, \btheta, F,\Delta \sim \mathrm{N} \left( \tilde{P}_1 \hat{\bbeta} - \tilde{P}_2 \btheta, \tilde{\mathcal{I}}_{\gamma \gamma}^{-1} \right),$$
as an approximation to $\bgamma| \by, \btheta, F,\Delta$.

Now consider taking expectation with respect to $\by \vert \btheta, D,\Delta$. Note that $\mathrm{E}_{\bgamma \vert \hat{\bbeta}, \btheta, F,\Delta} \left[ \lambda^*(\bbeta, \hat{\bbeta}, \Delta)\right]$ only depends on $\by$ through $\hat{\bbeta}$, thus it suffices to take expectation with respect to $\hat{\bbeta} \vert \btheta, D,\Delta$. 

Assumption of conditions A1, A3, A4, A6-A8, and by, for example, Theorem 4.1.3 of \citet[][page 111]{amemiya1985},
$$\hat{\bbeta}\vert \btheta, D, \Delta \sim \mathrm{N} \left( \tilde{\bbeta}, \tilde{\mathcal{K}} \right),$$
thus completing the proof.

LAN is a general property. However it may be non-trivial to verify for a given fitted likelihood. Instead, \citet{miller2019} provides the following conditions which are more restrictive but may be easier to verify. 

\begin{enumerate}
\item[B1]
$\hat{\bbeta}\in \mathcal{B}$ uniquely maximises $\log \pi (\by \vert \bt, F, \Delta)$.
\item[B2]
$\log \pi (\by \vert \bt, F, \Delta)$ has continuous third derivatives, uniformly bounded on $\mathcal{B}$.
\item[B3]
$\tilde{\mathcal{I}}$ is positive-definite.
\end{enumerate}

Assumption of B1-B3, along with A1, A3, A4 and A6, and by Theorem 3.2 of \cite{miller2019}, the fitted posterior converges to (5) in total variation.

\section{Derivation of expressions in Table~2} \label{sec:tab2proof}

\subsection{Self-information loss}

For the generator self information loss,
\begin{eqnarray}
\ell^*_{G,SI}(\btheta, \Delta) & = & \mathrm{E}_{\hat{\bbeta} \vert \btheta, D, \Delta} \left[  \mathrm{E}_{\bgamma \vert \hat{\bbeta}, \btheta, F, \Delta} \left[ N_{SI} - \frac{1}{2} \log \vert \tilde{\mathcal{I}} \vert + \frac{1}{2} \Vert \bbeta - \hat{\bbeta} \Vert^2_{2,\tilde{\mathcal{I}}} \right] \right] \nonumber \\
& = & N_{SI} - \frac{1}{2} \log \vert \tilde{\mathcal{I}} \vert \nonumber \\
& & \qquad \mbox{} + \frac{1}{2} \mathrm{E}_{\hat{\bbeta} \vert \btheta, D, \Delta} \left[  \mathrm{E}_{\bgamma \vert \hat{\bbeta}, \btheta, F, \Delta} \left[\Vert \bbeta - \hat{\bbeta} \Vert^2_{2,\tilde{\mathcal{I}}} \right] \right]. \label{eqn:si0}
\end{eqnarray}
The inner expectation on the right hand side of (\ref{eqn:si0}) is
\begin{eqnarray}
\mathrm{E}_{\bgamma \vert \hat{\bbeta}, \btheta, F, \Delta} \left[\Vert \bbeta - \hat{\bbeta} \Vert^2_{2,\tilde{\mathcal{I}}} \right] & = & \mathrm{E}_{\bgamma \vert \hat{\bbeta}, \btheta, F, \Delta} \left[ \Vert \bgamma - \hat{\bgamma} \Vert^2_{2,\tilde{\mathcal{I}}_{\gamma \gamma}} + 2 \left(\bgamma - \hat{\bgamma} \right)^\T \tilde{\mathcal{I}}_{\gamma \theta} \left(\btheta - \hat{\btheta} \right) \right. \nonumber \\
& & \qquad \mbox{} \left. + \Vert \btheta - \hat{\btheta} \Vert^2_{2,\tilde{\mathcal{I}}_{\theta \theta}}\right] \nonumber \\
& = & \Vert \tilde{P}_1 \hat{\bbeta} - P_2 \btheta - \hat{\bgamma} \Vert^2_{2, \tilde{\mathcal{I}}_{\gamma \gamma}} + p_{\gamma} \nonumber \\
& & + 2 \left(\tilde{P}_1 \hat{\bbeta} - P_2 \btheta - \hat{\bgamma}\right)^\T \tilde{\mathcal{I}}_{\gamma \theta} \left(\btheta - \hat{\btheta} \right) + \Vert \btheta - \hat{\btheta} \Vert^2_{2,\tilde{\mathcal{I}}_{\theta \theta}} \nonumber \\
& = & \Vert \btheta - \hat{\btheta} \Vert_{2,\tilde{\mathcal{T}}^{-1}_{\theta \theta}}^2 + p_{\gamma}. \label{eqn:si1}
\end{eqnarray}
Substituting (\ref{eqn:si1}) into (\ref{eqn:si0}) we obtain
\begin{eqnarray*}
\ell^*_{G,SI}(\btheta, \Delta) & = & N_{SI} - \frac{1}{2} \log \vert \tilde{\mathcal{I}} \vert + \frac{1}{2} \mathrm{E}_{\hat{\bbeta} \vert \btheta, D, \Delta} \left[ \Vert \btheta - \hat{\btheta} \Vert^2_{2,\tilde{\mathcal{T}}^{-1}_{\theta \theta}} + p_{\gamma} \right] \\
& = & N_{SI}- \frac{1}{2}\log \vert \tilde{\mathcal{I}} \vert + \frac{1}{2} \Vert \btheta - \tilde{\btheta} \Vert_{2,\tilde{\mathcal{T}}^{-1}_{\theta\theta}}^2 + \frac{1}{2} \mathrm{tr}\left( \tilde{\mathcal{T}}^{-1}_{\theta\theta}\tilde{\mathcal{K}}_{\theta\theta}\right) +p_\gamma/2
\end{eqnarray*}
as required.

\subsection{Squared error loss}

For the generator squared error loss
$$\lambda^*_{G,SE}(\bbeta, \hat{\bbeta}, \Delta) = \Vert \bbeta - \hat{\bbeta} \Vert^2_{2,I_p},$$
and
\begin{equation}
\ell^*_{G,SE}(\btheta, \Delta) = \mathrm{E}_{\hat{\bbeta} \vert \btheta, D, \Delta} \left[  \mathrm{E}_{\bgamma \vert \hat{\bbeta}, \btheta, F, \Delta} \left[ \Vert \bbeta - \hat{\bbeta} \Vert^2_{2,I_p} \right] \right].
\label{eqn:se0}
\end{equation}
The inner expectation is
\begin{eqnarray}
\mathrm{E}_{\bgamma \vert \hat{\bbeta}, \btheta, F, \Delta} \left[ \Vert \bbeta - \hat{\bbeta} \Vert^2_{2,I_p} \right] & = & \mathrm{E}_{\bgamma \vert \hat{\bbeta}, \btheta, F, \Delta} \left[ \Vert \bgamma - \hat{\bgamma} \Vert^2_{2,I_{p_\gamma}} + \Vert \btheta - \hat{\btheta} \Vert^2_{2,I_{p_\theta}} \right] \nonumber \\
& = & \Vert \tilde{P}_1\hat{\bbeta} - \tilde{P}_2 \btheta - \hat{\bgamma} \Vert^2_{2,I_{p_\gamma}} + \mathrm{tr}\left( \tilde{\mathcal{I}}_{\gamma \gamma}^{-1} \right) + \Vert \btheta - \hat{\btheta} \Vert^2_{2,I_{p_\theta}}\nonumber \\
& = & \Vert \btheta - \hat{\btheta} \Vert^2_{2,\tilde{\mathcal{S}}_{\theta \theta}} + \mathrm{tr}\left( \tilde{\mathcal{I}}_{\gamma \gamma}^{-1} \right).
\label{eqn:se1}
\end{eqnarray}
Substituting (\ref{eqn:se1}) into (\ref{eqn:se0}) we obtain
\begin{eqnarray*}
\ell^*_{G,SE}(\btheta, \Delta) &=& \mathrm{E}_{\hat{\bbeta} \vert \btheta, D, \Delta} \left[\Vert \btheta - \hat{\btheta} \Vert^2_{2,\tilde{\mathcal{S}}_{\theta \theta}} + \mathrm{tr}\left( \tilde{\mathcal{I}}_{\gamma \gamma}^{-1} \right) \right] \\
& = & \Vert \btheta - \tilde{\btheta} \Vert^2_{2,\tilde{\mathcal{S}}_{\theta \theta}} + \mathrm{tr}\left( \tilde{\mathcal{I}}_{\gamma \gamma}^{-1} \right) + \mathrm{tr}\left(\tilde{\mathcal{S}}_{\theta \theta} \tilde{\mathcal{K}}_{\theta \theta} \right)\\
& = & \Vert \btheta - \tilde{\btheta} \Vert^2_{2,\tilde{\mathcal{S}}_{\theta \theta}} + \mathrm{tr}\left( \tilde{\mathcal{I}}^{-1} \right) + \mathrm{tr}\left[\tilde{\mathcal{S}}_{\theta \theta} \left(\tilde{\mathcal{K}}_{\theta \theta}-\tilde{\mathcal{T}}_{\theta \theta} \right)\right]
\end{eqnarray*}
as required. The last line follows from
$$\mathrm{tr}\left( \tilde{\mathcal{I}}^{-1} \right) = \mathrm{tr}\left( \tilde{\mathcal{I}}_{\gamma \gamma}^{-1} + \tilde{\mathcal{I}}_{\gamma \gamma}^{-1} \tilde{\mathcal{I}}_{\gamma \theta}\tilde{\mathcal{T}}_{\theta \theta}\tilde{\mathcal{I}}_{\theta \gamma}\tilde{\mathcal{I}}_{\gamma \gamma}^{-1} \right) + \mathrm{tr} \left(\tilde{\mathcal{T}}_{\theta \theta}\right).$$

\subsection{Entropy loss}

The composite entropy loss is
\begin{eqnarray*}
\ell^*_{C,E}(\btheta, \Delta) & = & - \mathrm{E}_{\bbeta \vert \by, F, \Delta} \left[ \log \pi (\bbeta \vert \by, F, \Delta) \right]\\
& = & N_{SI} - \frac{1}{2} \log \vert \tilde{\mathcal{I}} \vert + \frac{p}{2}
\end{eqnarray*}
under (5).

\subsection{Trace variance loss}

The composite trace variance loss is
\begin{eqnarray*}
\ell^*_{C,TV}(\btheta, \Delta) & = & \mathrm{tr} \left( \mathrm{var} \left( \bbeta \vert \by, F, \Delta \right) \right) \\
& = & \mathrm{tr} \left(  \tilde{\mathcal{I}}^{-1} \right)
\end{eqnarray*}
under (5).

\section{Derivation of expressions (7) and (8) in Section~4.1} \label{sec:67}

The self-information loss function is given by
\begin{eqnarray*}
\lambda_{SI}(\bgamma, \by, F, \Delta) & = & - \log \pi(\bgamma \vert \by, F, \Delta)\\
& = & \log \Gamma \left( \frac{\hat{a}_F}{2} \right) + \frac{p}{2} \log \pi - \log \Gamma \left( \frac{\hat{a}_F+p}{2} \right) + \frac{p}{2} \log \hat{b}_F + \frac{1}{2} \log \vert \hat{V}_F \vert \\
& & + \frac{\hat{a}_F + p}{2} \log \left( 1 + \frac{1}{\hat{a}_F} \left( \bgamma - \hat{\bmu}_F \right)^\T \left( \frac{\hat{b}_F}{\hat{a}_F} \hat{V}_F \right)^{-1} \left( \bgamma - \hat{\bmu}_F \right) \right),
\end{eqnarray*}
where $\Gamma(\cdot)$ is the Gamma function. The expected value of $\lambda_{SI}(\bgamma, \by, F, \Delta)$ with respect to the fitted posterior distribution of $\bgamma$ is
\begin{eqnarray}
\mathrm{E}_{\bgamma \vert \by, F, \Delta} \left[ \lambda_{SI}(\bgamma, \by, F, \Delta)\right] & = & \log \Gamma \left( \frac{\hat{a}_F}{2} \right) + \frac{p}{2} \log \pi - \log \Gamma \left( \frac{\hat{a}_F+p}{2} \right) \nonumber \\
& & + \frac{p}{2} \log \hat{b}_F + \frac{1}{2} \log \vert \hat{V}_F \vert  \nonumber \\
& & +\frac{\hat{a}_F + p}{2} \left( \psi \left( \frac{\hat{a}_F+p}{2} \right) - \psi \left( \frac{p}{2} \right) \right) \label{eqn:60}\\
& = & H_{SI,1} + \frac{1}{2} \log \vert \hat{V}_F \vert + \frac{p}{2} \log \hat{b}_F, \label{eqn:61}
\end{eqnarray}
where (\ref{eqn:60}) follows from, for example, \citet[][page 23]{kotznadarajah2004} with $\psi(\cdot)$ the digamma function. The expression (7) is given by taking expectation of (\ref{eqn:61}) with respect to the marginal distribution of $\by$ under the designer model.

The squared error loss is
$$\lambda_{SI}(\bgamma, \by, F, \Delta) = \Vert \bgamma - \hat{\bmu}_F \Vert^2_{2,I_p}.$$
From properties of the multivariate t-distribution \citep[for example][page 11]{kotznadarajah2004}, the expected value of $\lambda_{SE}(\bgamma, \by, F, \Delta)$ with respect to the fitted posterior distribution of $\bgamma$ is
\begin{eqnarray}
\mathrm{E}_{\bgamma \vert \by, F, \Delta} \left[ \lambda_{SE}(\bgamma, \by, F, \Delta)\right] & = & \mathrm{tr} \left( \mathrm{var} \left( \bgamma \vert \by, F, \Delta \right)\right) \nonumber \\
& = & \frac{\hat{b}_F}{\hat{a}_F-2} \mathrm{tr} \left( \hat{V}_F \right). \label{eqn:70}
\end{eqnarray}
The expression (8) is given by taking expectation of (\ref{eqn:70}) with respect to the marginal distribution of $\by$ under the designer model where
\begin{eqnarray*}
\mathrm{E}_{\by \vert D, \Delta} \left[\hat{b}_F\right] & = & \mathrm{E}_{\by \vert D, \Delta} \left[ b_F + \Vert \by - X \bmu_F \Vert^2_{2,\Sigma_F^{-1}}\right]\\
& = & b_F + \Vert \bm_D - X\bmu_F \Vert^2_{2,\Sigma_F^{-1}} + \mathrm{tr} \left(\Sigma_D \Sigma_F^{-1} \right).
\end{eqnarray*}

\section{Derivation of expressions (9) and (10) in Section~4.1} \label{sec:89}

Under the fitted prior distribution with $V_F^{-1} = 0_{p \times p}$,
\begin{eqnarray*}
\Sigma_F^{-1} & = & \left(I_n + X V_F X^\T \right)^{-1}\\
& = & I_n - X\left(X^\T X \right)^{-1} X^\T\\
& = & I_n - H_X
\end{eqnarray*}
where $H_X = X\left(X^\T X \right)^{-1} X^\T$. Under the designer prior distribution with $V_D = \kappa \left(Z^\T Z\right)^{-1}$,
$$\Sigma_D = \frac{b_D}{a_D-2} \left(I_n + \kappa H_Z \right),$$
where $H_Z = Z\left(Z^\T Z \right)^{-1} Z^\T$. Then with $b_F=0$, $\bmu_F = \bzero_p$ and $\bmu_D = \bzero_q$
\begin{eqnarray*}
\mathrm{E}_{\by \vert D, \Delta} \left( \hat{b}_F \right) & = & b_F + \mathrm{tr} \left( \Sigma^{-1}_F\Sigma_D \right)\\
& = & \frac{b_D}{a_D-2} \mathrm{tr} \left[ \left(I_n - H_X \right)\left(I_n + \kappa H_Z\right)\right]\\
& = & \frac{b_D}{a_D-2} \mathrm{tr} \left( I_n + \kappa H_Z - H_X - \kappa H_XH_Z \right)\\
& = & \frac{b_D}{a_D-2} \left[ n + \kappa q - p - \kappa \mathrm{tr}\left(H_XH_Z \right)\right]\\
& = & \frac{b_D}{a_D-2} \left[ (1+\kappa)n - p - \kappa \mathrm{tr}\left(H_XH_Z \right) - \kappa d\right]
\end{eqnarray*}
with the last line following from $d=n-q$. Now
$$\mathrm{tr}\left(H_XH_Z \right) = \mathrm{tr}\left[H_X Z \left(Z^\T Z\right)^{-1} Z^\T\right] = \mathrm{tr}\left[Z^\T H_X Z \left(Z^\T Z\right)^{-1} \right].$$
The $q \times q$ matrix $Z^\T Z$ is diagonal with the $l$th diagonal element, denoted $r_l$, giving the number of times treatment $l=1,...,q$ is replicated. Note that $\sum_{l=1}^q r_l = n$. Since $Z^\T Z$ is diagonal,
$$\mathrm{tr}\left(H_XH_Z \right) = \sum_{l=1}^q \frac{1}{r_l} \left[ Z^\T H_X Z\right]_{ll},$$
where $\left[ Z^\T H_X Z\right]_{ll}$ is the $l$th diagonal element of $Z^\T H_X Z$. Now
$$\left[ Z^\T H_X Z\right]_{ll} = \sum_{i=1}^n \sum_{j=1}^n z_{il}z_{jl} h_{Xij},$$
with $h_{Xij}$ the $ij$th element of $H_X$ and $z_{il}$ the $il$th element of $Z$. Note, if runs $i$ and $j$ receive the same treatment (say $l$), then $h_{Xij} = h_{Xii} = h_{Xjj}$ and $z_{il} = z_{jl} = 1$. Then
\begin{eqnarray*}
\mathrm{tr}\left(H_XH_Z \right) &=& \sum_{l=1}^q \frac{1}{r_l} \sum_{i=1}^n \sum_{j=1}^n z_{il}z_{jl} h_{Xij}\\
& = & \sum_{l=1}^q \frac{1}{r_l} \sum_{i=1}^n z_{il} \sum_{j=1}^n z_{jl} h_{Xij}\\
& = & \sum_{l=1}^q \frac{1}{r_l} \sum_{i=1}^n z_{il} r_l h_{Xii}\\
& = & \sum_{l=1}^q \sum_{i=1}^n z_{il} h_{Xii}\\
& = & \sum_{i=1}^n h_{Xii} \sum_{l=1}^q z_{il} \\
& = & \sum_{i=1}^n h_{Xii} \\
& = & p.
\end{eqnarray*}
Then
\begin{eqnarray}
\mathrm{E}_{\by \vert D, \Delta} \left( \hat{b}_F \right) &=&\frac{b_D}{a_D-2} \left[ (1+\kappa)n - p - \kappa p - \kappa d\right] \nonumber \\
& = & \frac{b_D}{a_D-2} \left[ (1+\kappa)(n - p) - \kappa d\right] \label{eqn:last}
\end{eqnarray}
Lastly (\ref{eqn:last}) is substituted into expressions (7) and (8) ignoring constants that do not depend on $\Delta$.

\section{Estimating parameters of cubic spline models} \label{sec:cubic}

Suppose the aim is to learn the relationship between a scalar controllable variable $x \in \mathcal{X} = [0,1]$ and a response $y$. For $i=1,\dots,n$, it is assumed that
$$y_i \sim \mathrm{N}\left( \mu(x_i), \sigma^2 \right),$$
where $\mu(x)$ and $\sigma^2$ are unknown. The function $\mu(x)$ can be approximated by the Michaelis-Menten model given by (13). However, this model is known to be an incomplete representation of reality. In light of this, the fitted model is assumed to have
$$\mu(x) = \sum_{j=1}^m \gamma^{(m)}_j g^{(m)}_j(x),$$
where $\left\{g^{(m)}_1(x),\dots, g^{(m)}_m(x) \right\}$ are a set of $m$ basis functions and $\bgamma^{(m)} = \left(\gamma_1^{(m)},\dots,\gamma_m^{(m)}\right)^\T$ are unknown parameters. We let $\left\{g^{(m)}_1(x),\dots, g^{(m)}_m(x) \right\}$ be cubic spline basis functions \citep[for example][Chapter 5]{woods2017}. Instead of fixing $m$ we allow $m$ to be unknown with the constraint $4 \le m \le n$ (the lower bound following from the assumption of cubic splines).

The design problem is to choose the values of $\Delta = \left(x_1,\dots,x_n\right)^\T$.

The fitted prior distributions for $\bgamma^{(m)}$ and $\sigma^2$ are $\bgamma^{(m)} \vert \sigma^2, F, \Delta \sim \mathrm{N}\left( \bzero_m, \sigma^2 \kappa I_m \right)$ and $\sigma^2 \vert F, \Delta \sim \mathrm{IG}\left(a_F/2, b_F/2 \right)$, where $\kappa$ is known. The fitted prior distribution for the discrete number of basis functions, $m$, is such that $\pi(m) = 1/(n-5)$, i.e. a uniform prior.

The model-averaged fitted posterior mean of $\mu(x)$ is
\begin{equation}
\mathrm{E}\left( \mu(x) \vert \by, F, \Delta \right) = \sum_{m=4}^n \left[\pi(m \vert \by, F, \Delta) \sum_{j=1}^m g_j(x) \mathrm{E}\left( \gamma_j^{(m)} \vert \by, m, F, \Delta \right)\right].
\label{eqn:pmmean}
\end{equation}
In (\ref{eqn:pmmean}), $\mathrm{E}\left( \gamma_j^{(m)} \vert \by, m, F, \Delta \right)$ is the $j$th element of the fitted posterior mean of $\bgamma^{(m)}$ (conditional on $m$ basis functions, for $j=1,\dots,m$) given by
$$\mathrm{E}\left( \bgamma^{(m)} \vert \by, m, F, \Delta \right) = \left( \kappa^{-1} I_m + G_m^\T G_m \right)^{-1} G_m^\T \by,$$
where $G_m$ is the $n \times m$ matrix with $j$th column $\left(g^{(m)}_j(x_1), \dots, g^{(m)}_j(x_n) \right)^\T$, for $j=1,\dots,m$. Additionally, $\pi(m \vert \by, F, \Delta)$ is the fitted posterior probability of $m$ given by
\begin{equation}
\pi(m \vert \by, F, \Delta) \propto \vert R_m \vert^{-\frac{1}{2}} \left( b_F + \Vert \by \Vert^2_{2, R_m} \right)^{- \frac{a_F + n}{2}},
\label{eqn:pmprob}
\end{equation}
where $R_m = I_n + \kappa G_mG_m^\T$. The expressions (\ref{eqn:pmmean}) and (\ref{eqn:pmprob}) follow from the fact that the fitted model is a normal linear model and can be found in, for example, \citet[][Chapter 11]{ohaganforster2004}.

The chosen loss function is predictive squared error 
\begin{equation}
\lambda(\by, \mu(\cdot) , F, \Delta) = \int_{\mathcal{X}} \left( \mu(x) - \mathrm{E}\left( \mu(x) \vert \by, F, \Delta \right) \right)^2 \mathrm{d}x,
\label{eqn:lossy}
\end{equation}
a measure of the accuracy of the model-averaged fitted posterior mean of $\mu(x)$.

We consider finding designs by minimising
\begin{enumerate}
\item[(a)]
internal expected predictive squared error loss; and
\item[(b)]
external expected predictive squared error loss.
\end{enumerate}
For (b) external expected predictive squared error loss, the disjoint designer model has $\mu(x) = \eta(\bxi, x)$ where $\eta(\cdot, x)$ is given by (13) with independent designer prior distributions of $\xi_1 \sim \mathrm{U}[20,200]$, $\xi_2 \sim \mathrm{U}[20,200]$ and $\sigma^2 \sim \mathrm{IG}\left(a_D/2, b_D/2\right)$. 

\begin{center}
\begin{figure}
\caption{Designs for the cubic spline model showing the location of $x$ for each of the two designs, i.e. found by minimising (a) external and (b) internal expected predictive squared error loss. The size of the plotting symbol is an increasing function of the replication.} \label{fig:cubic} 
\includegraphics[scale=0.475]{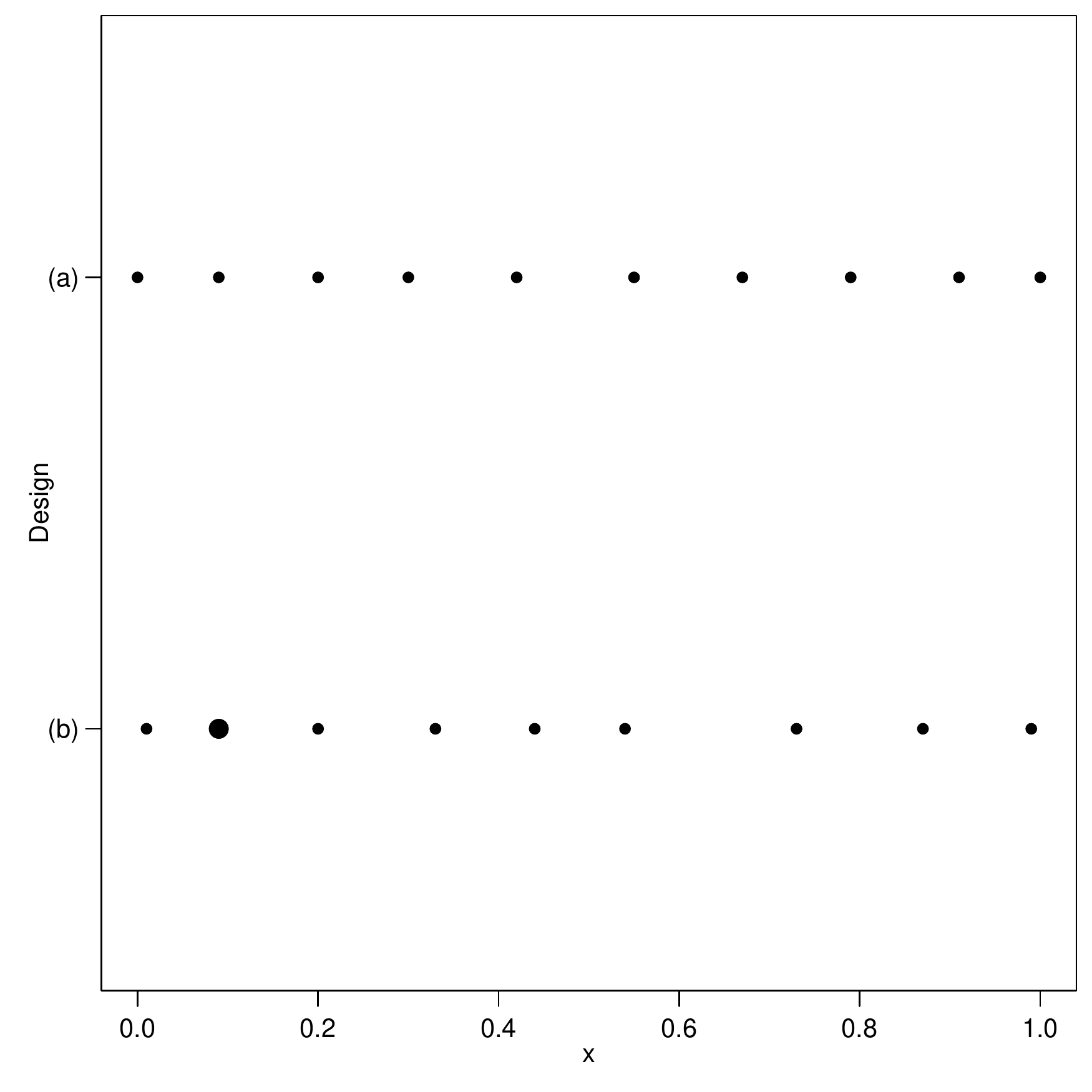}
\end{figure}
\end{center}

Neither of the expected loss functions are available in closed form and require approximation for which we use Monte Carlo. The internal expected predictive squared error loss can be approximated using the following steps.
\begin{enumerate}
\item
For $b=1, \dots, B$, generate $m_b$ uniformly from $\left\{4, \dots, n \right\}$, $\sigma_b^2$ from $\mathrm{IG}\left(a_F/2,b_F/2\right)$, $\bgamma^{(m_b)}$ from $\mathrm{N}\left(\bzero_{m_b}, \sigma_b^2 \kappa I_{m_b} \right)$. Then generate $\by_b$ from $\mathrm{N}\left(G_{m_b} \bgamma^{(m_b)}, \sigma_b^2 I_{n} \right)$.
\item
Approximate the internal expected predictive squared error loss by
$$\hat{L}_{FF,PSE}(\Delta) = \frac{1}{B}\sum_{b=1}^B \lambda\left[\by_b, \sum_{j=1}^{m_b} g_j^{(m_b)}(\cdot) \gamma_{bj}^{(m_b)} , F, \Delta\right]$$
where $\gamma_{bj}^{(m_b)}$ is the $j$th element of $\bgamma^{(m_b)}$.
\end{enumerate}
The external expected predictive squared error loss can be approximated using the following steps.
\begin{enumerate}
\item
For $b=1, \dots, B$, generate $\bxi_b$ from the designer prior distribution and $\sigma_b^2$ from $\mathrm{IG}\left(\frac{a_D}{2}, \frac{b_D}{2}\right)$.Then generate $\by_b$ from $\mathrm{N}\left(\bseta(\bxi_b, \Delta), \sigma_b^2 I_{n} \right)$.
\item
Approximate the internal expected predictive squared error loss by
$$\hat{L}_{DF,PSE}(\Delta) = \frac{1}{B}\sum_{b=1}^B \lambda \left[\by_b, \eta(\bxi_b, \cdot) , F, \Delta \right].$$
\end{enumerate}
For both approximations, the integration in (\ref{eqn:lossy}) is evaluated using quadrature.

The hyperparameters of the fitted and designer prior distributions of $\sigma^2$ are set as $a_F=a_D = 6$ and $b_F = b_D = 4$, ensuring that the prior mean and variance are both one. The value of $\kappa$ is chosen to be $10^6$, indicating a very vague fitted prior distribution for $\bgamma^{(m)}$.
Designs are found with $n=10$ using the approximate coordinate exchange algorithm.

Table~\ref{tab:cubic} shows the efficiencies of the two designs under internal and external predictive squared error loss. Figure~\ref{fig:cubic} shows the values of $\Delta = \left(x_1,\dots,x_n\right)^\T$ for the two designs (where the size of the plotting symbol is an increasing function of the replication.). The design found by minimising internal expected loss is similar to an evenly-spaced design whereas the design found by minimising external expected loss appears to be a compromise between the evenly-spaced design and the structure of points found for the Michaelis-Menten model in Section~4.2. Further, it was found that the internal expected loss was flat near the design minimising its value, evidenced by the high efficiency under the internal expected loss for the design minimising the external expected loss. This agrees with the observation of \citet{ryanetal2016} that expected loss functions become flat under vague prior distributions (i.e. in this example $\kappa = 10^6$). By contrast, the design found by minimising the external expected loss provides a design of high efficiency for the fitted model but reflects best current knowledge of the underlying process.

\begin{table}[t]
\begin{center}
\caption{Efficiencies for the two designs found for the cubic spline model. \label{tab:cubic}}

\vspace{2mm}

\begin{tabular}{lrr} \hline \hline
 & \multicolumn{2}{c}{Efficiency} \\ \hline
 & Internal PSE (a) & External PSE (b) \\ \hline
Internal PSE (a) & 100 & 84.0 \\
External PSE (b) & 97.1 & 100 \\ \hline \hline
\end{tabular}
\end{center}
\end{table}

\bibliographystyle{rss}

\bibliography{BayesDesignAlt}
\end{document}